\documentclass[11pt]{article}

\usepackage{fullpage}
\usepackage[utf8]{inputenc}

\usepackage{graphicx}
\usepackage{amsmath,amsthm,amssymb}
\usepackage{algorithm}
\usepackage{algpseudocode}
\usepackage{multirow}
\usepackage{makecell}

\usepackage{thm-restate,color,xspace}
\usepackage{comment}
\usepackage{thmtools}
\usepackage{xcolor}
\usepackage{nameref}
\usepackage{array}

\definecolor{ForestGreen}{rgb}{0.1333,0.5451,0.1333}
\definecolor{DarkGreen}{RGB}{0,100,0}
\definecolor{DarkRed}{rgb}{0.65,0,0}
\definecolor{Red}{rgb}{1,0,0}
\usepackage[linktocpage=true,
pagebackref=true,colorlinks,
linkcolor=DarkRed,citecolor=ForestGreen,
bookmarks,bookmarksopen,bookmarksnumbered]
{hyperref}
\usepackage[capitalize]{cleveref}

\declaretheorem[numberwithin=section]{theorem}
\declaretheorem[numberlike=theorem]{lemma}

\declaretheorem[numberlike=theorem]{proposition}

\declaretheorem[numberlike=theorem]{corollary}

\declaretheorem[numberlike=theorem]{claim}

\declaretheorem[numberlike=theorem]{assumption}
\declaretheorem[numberlike=theorem]{observation}

\declaretheorem[numberlike=theorem,style=definition]{definition}
\declaretheorem[numberlike=theorem,style=definition]{remark}

\global\long\def\OPT{\mathrm{OPT}}
\global\long\def\poly{\mathrm{poly}}

\global\long\def\dist{\mathrm{dist}}

\usepackage{bbm}

\usepackage{tikz}
\usetikzlibrary{calc,positioning,decorations.pathreplacing,arrows.meta,shapes,fit,backgrounds}

\newcommand{\DSN}{\textnormal{\textsc{DSN}}\xspace}
\newcommand{\polylog}{\operatorname{polylog}}

\newcommand{\Dcal}{\mathcal{D}}
\newcommand{\Ccal}{\mathcal{C}}
\newcommand{\Tcal}{\mathcal{T}}
\newcommand{\HS}{\mathsf{HS}}
\newcommand{\LP}{\mathsf{LP}}
\newcommand{\VCA}{\mathsf{VCA}}
\newcommand{\sh}{\mathsf{sh}}
\newcommand{\SF}{\mathsf{SF}}
\newcommand{\xbf}{\mathbf{x}}
\newcommand{\cbf}{\mathbf{c}}
\newcommand{\abf}{\mathbf{a}}
\newcommand{\Uniform}{\mathsf{Uniform}}

\newcommand{\paperopenquestions}{\section{Open Questions}
\label{sec:conclusion}

We conclude by highlighting several interesting directions that our work leaves open.

\begin{enumerate}
    \item Can we augment connectivity by more than one? Even augmenting by two is not handled by our current techniques.
    \item The eventual goal, and a central open problem, is to obtain a polynomial-time $\poly(k,\log n)$-competitive online algorithm for general VC-SNDP.
    \item Derandomization: Can we obtain a deterministic, polynomial-time $O((k+\log n)\log n)$-competitive algorithm for OVCA?
     Can we obtain a deterministic, polynomial-time $O(k+\log n)$-approximation algorithm for VCA?
\end{enumerate}
}

\newcommand{\paperacknowledgements}{\section*{Acknowledgements}

\paragraph{Disclosure of Generative AI Use.} For the results and techniques, the deterministic OVCA algorithm and all techniques behind it (\Cref{ssec:component-cuts} to \Cref{sect:DeterministicOVCA}, and \Cref{sec:necessity}) were developed by the authors without AI assistance. This part presents the main technical contributions of this paper. The online $\{1,2\}$-VC-SNDP algorithm and the corresponding techniques (\Cref{sec:1-2-online}) were also developed by the authors. The polynomial-time randomized online and offline VCA algorithms (\Cref{sect:RandomizedOVCA} and \Cref{sect:OfflineVCA}) were developed with assistance from AI tools including OpenAI GPT 5.5 and Claude Opus 4.8 (the technique is to replace part of the deterministic algorithm with a standard online covering LP solver and an online rounding). The hardness results (\Cref{sec:hardness}) were discovered with assistance from the AI tools above.

For writing, an initial draft was created using AI tools (Claude Opus 4.8). The main body was then rewritten by the authors.
}

\crefname{fact}{Fact}{Facts}
\Crefname{fact}{Fact}{Facts}
\crefname{example}{Example}{Examples}
\Crefname{example}{Example}{Examples}

\title{Online Vertex Connectivity Augmentation}
\author{Yaowei Long\thanks{Part of the work was done while the author was an intern at Microsoft Research.}\\
\normalsize University of Michigan\\
\normalsize\texttt{yaoweil@umich.edu}
\and Sepideh Mahabadi\\
\normalsize Microsoft Research\\
\normalsize\texttt{smahabadi@microsoft.com}
\and Jakub Tarnawski\\
\normalsize Microsoft Research\\
\normalsize\texttt{jatarnaw@microsoft.com}}
\date{}

\begin{document}
\maketitle
\pagenumbering{gobble}

\begin{abstract}
Network design problems form a classic subfield of online algorithms, but almost nothing is known about their more challenging vertex-connectivity variants. We initiate the study of online vertex-connectivity network design for arbitrary demand pairs by considering the fundamental problem of connectivity augmentation.

In an instance of \emph{Online Vertex Connectivity Augmentation} (OVCA), we are given a $k$-vertex-connected graph $F$ and a set of additional purchasable edges (with costs). Demand pairs $(s_1,t_1),(s_2,t_2),\dots$ then arrive online. Upon each arrival, the algorithm must irrevocably purchase edges so that $s_i$ and $t_i$ become $(k{+}1)$-vertex-connected in the union of $F$ with all edges purchased so far.

Our main contribution is an $O((k+\log n)\log n)$-competitive online algorithm for OVCA. We also give an offline $O(k+\log n)$-approximation algorithm. We obtain our algorithms via a structural dichotomy on the minimum vertex cuts of $F$.

We note that some dependence on $k$ in the approximation ratio is necessary -- i.e., the problem is hard to approximate within any $\polylog(n)$ factor independent of $k$ -- which shows a separation from the edge-connectivity setting.

Finally, specializing to $k=1$, our methods yield the first polylogarithmic-competitive online algorithm for a vertex-connectivity survivable network design problem, namely an $\widetilde O(\log^3 n)$-competitive algorithm for online $\{1,2\}$-VC-SNDP (i.e., an SNDP variant where each demand pair comes with a vertex-connectivity requirement of $1$ or $2$).
\end{abstract}

\newpage

\tableofcontents
\newpage
\pagenumbering{arabic}

\section{Introduction}
\label{sec:intro}

The \emph{Survivable Network Design Problem} (SNDP) is a classic graph problem that asks for a minimum-cost subgraph meeting prescribed pairwise (edge- or \mbox{vertex-}) connectivity requirements.  Namely, given an edge-weighted graph $G=(V,E)$ and requirements $\{r_{u,v}\colon u,v\in V, u\neq v\}$, the subgraph must include at least $r_{u,v}$ edge-disjoint or internally vertex-disjoint paths between $u$ and $v$.
SNDP generalizes several important problems including the polynomially solvable minimum
spanning tree (MST), and $s$-$t$ shortest path, as well as several NP-hard problems, including Steiner Tree and Steiner Forest~\cite{Jain01,WilliamsonGMV95}. A special case of interest is the {\em connectivity augmentation problem} where a partial solution is given for free that already satisfies partial connectivity requirements.

SNDP has applications in a wide variety of domains, and as a result the problem and its variants have been studied in many computational models. In this work we consider the \emph{online} model, where the connectivity requirements arrive one at a time and the algorithm must purchase edges irrevocably to satisfy all the requirements that have arrived so far.
Here the two basic flavors of SNDP diverge dramatically.

For \emph{edge}-connectivity SNDP, Gupta, Krishnaswamy, and Ravi~\cite{GuptaKR09} gave an $O(k\log^3 n)$-competitive randomized algorithm, where $k=\max_{u,v}r_{u,v}$ is the largest requirement, and this has been the state of the art for over fifteen years.
For \emph{vertex}-connectivity SNDP the picture is far bleaker.
Even offline, the best known approximation is $O(k^3\log n)$ due to Chuzhoy and Khanna~\cite{ChuzhoyK09}, and strong hardness is known~\cite{KortsarzKL04,Laekhanukit14,Manurangsi19} -- in particular,
even in the single-source setting, there is a $k^{\Omega(1)}$ hardness of approximation~\cite{chakraborty2008network}.
Online, essentially nothing was known, and the sole prior result is the bicriteria single-source algorithm of Naor, Panigrahi, and Singh~\cite{NaorPS11}.
General online vertex-connectivity SNDP has remained wide open.
There are structural reasons for this. The standard edge-cut uncrossing arguments do not directly apply to arbitrary-pair vertex-connectivity augmentation. Moreover, a single vertex cut can induce many covering constraints that a solution must satisfy.

\paragraph{The augmentation problem.}
We attack the difficulty in its most basic incremental form.
Suppose a network is already $k$-vertex-connected and we wish to make selected pairs $(k{+}1)$-vertex-connected by purchasing additional links as demands arrive.
This is the natural online variant of vertex-connectivity {augmentation}, a problem with a rich offline history~\cite{Frank92,JacksonJ05,VeghB08,Vegh11,CheriyanT99,KortsarzN08,Nutov11SubsetAug,Nutov24}.
We formalize it here.

\begin{definition}[OVCA]
\label{def:problem}
An instance of \emph{Online Vertex Connectivity Augmentation} consists of an integer $k\ge 1$; a weighted undirected graph $G=(V,E,c)$ with $c:E\to\mathbb R_{\ge0}$; and a spanning subgraph $F$ with $E(F)\subseteq E$, $c(e)=0$ for all $e\in E(F)$, such that $F$ is $k$-vertex-connected.
We call $F$ the \emph{free graph} and $U:=E\setminus E(F)$ the \emph{candidate} (or \emph{purchasable}) edges.
A collection $\Dcal$ of demand pairs $(s_1,t_1),(s_2,t_2),\dots$ arrives online, and upon each arrival the algorithm irrevocably purchases a subset of~$U$.
We use $X\subseteq U$ for all edges purchased so far, and we must maintain that $s_i$ and $t_i$ are $(k{+}1)$-vertex-connected in $(V,E(F)\cup X)$ for every past demand~$i$.
The \emph{competitive ratio} is the worst-case ratio of $\sum_{e\in X}c(e)$ to the optimal offline cost.
\end{definition}

\paragraph{Results.}
Our main result is the first general online algorithm for vertex connectivity augmentation.

\begin{theorem}[Main result]
\label{thm:main-intro}
There are two online algorithms for OVCA:
\begin{enumerate}
\item a \emph{deterministic} algorithm with competitive ratio $O((k+\log n)\log n)$ and running time $\poly(2^k,n)$; in particular, for $k=O(\log n)$ it runs in polynomial time and is $O(\log^2 n)$-competitive;
\item a \emph{Las~Vegas randomized}\footnote{All randomized online algorithms in this paper assume an oblivious adversary. That is, the entire online sequence is fixed in advance, independently of the algorithm's random choices.} algorithm with expected competitive ratio $O((k+\log n)\log n)$ and polynomial expected running time.
\end{enumerate}
\end{theorem}

At the core of this result is a structural dichotomy on the minimum vertex cuts of~$F$.
We do not know how to handle all cuts in a unified way.
Instead, our crucial technical insight is that those minimum cuts whose removal leaves at least $k+1$ connected components, which we call \emph{shattering} cuts,
can all be handled
via a novel reduction to a single, shared node-weighted Steiner forest problem instance on an auxiliary graph.
Unfortunately,
the correctness of this construction breaks down for non-shattering cuts.
However, it turns out that
the non-shattering cuts
induce a number of covering constraints
that can be upper-bounded such that
they can all
be handled by an appropriate online hitting set subroutine.

The first
part of \Cref{thm:main-intro}
is a \emph{deterministic} algorithm (\Cref{thm:deterministic}) with running time $\poly(2^k,n)$, whose proof already exposes all our main structural ideas.
The second part removes the exponential dependence on~$k$ from the running time, without changing the competitive ratio, by maintaining
constraints
{fractionally} through an oracle-efficient online covering framework and rounding them online (\Cref{thm:randomizedOVCA}).

We also show that in the offline setting,
the approximation guarantee can be optimized further, as shown in \Cref{thm:offline-intro} (see \Cref{thm:OfflineVCA} for a detailed version).
The offline variant of OVCA will be called \emph{VCA}.

\begin{theorem}[Offline VCA]
\label{thm:offline-intro}
There is a Las Vegas randomized algorithm for VCA in the offline setting with approximation $O(k+\log n)$ and polynomial running time.
\end{theorem}

To put this in context, the best approximation ratio for VCA
that follows by applying prior known results
would be an $O(k^3\log n)$-approximation
via VC-SNDP~\cite{ChuzhoyK09}.

It is not difficult to see that OVCA generalizes the online Steiner forest problem already for $k=1$ (see \cref{prop:log-lb}),
and thus inherits its $\Omega(\log n)$ lower bound on the competitive ratio~\cite{ImaseW91}.
However, it is natural to ask whether the dependence on the connectivity parameter~$k$ in our ratios could be avoided.
It cannot: in the offline setting,
the (undirected) VCA problem
is, somewhat surprisingly,
no easier than the \emph{Directed} Steiner Network problem~\cite{LandoN09},
which does not admit $\polylog(n)$-approximations
(either offline~\cite{DodisK99} or online~\cite{HLMT27}).
This rules out polynomial-time $\polylog(n)$-approximation algorithms for VCA, as well as
any
$\polylog(n)$-competitive algorithms
for OVCA.

\begin{restatable}[Hardness]{theorem}{hardnessconsequencesthm}
\label{thm:hardness-intro}
Let $\epsilon>0$ be an arbitrary constant.
\begin{itemize}
\item Unless $\mathrm{NP}\subseteq\mathrm{DTIME}(n^{\polylog n})$, VCA has no deterministic polynomial-time algorithm with approximation $O(2^{\log^{1-\epsilon}n})$ independent of $k$.
\item Unless $\mathrm{NP}\subseteq\mathrm{BPTIME}(n^{\polylog n})$, VCA has no randomized polynomial-time algorithm with
approximation $O(2^{\log^{1-\epsilon}n})$ independent of $k$.
\item Unconditionally, the competitive ratio of any algorithm for OVCA
is at least $2^{\Omega(\sqrt{\log n})}$. In particular, OVCA has no $\polylog(n)$-competitive algorithm independent of $k$.
This holds even for randomized algorithms against an oblivious adversary.
\end{itemize}
\end{restatable}

Finally, the case $k=1$ is already interesting on its own: a $1$-connected free graph contains a spanning tree, and ``augment to $2$-vertex-connected'' is exactly the survivable-design requirement $r\in\{1,2\}$.
In this case, we can produce the free graph required by the OVCA problem assumption using
random low-stretch spanning trees~\cite{ElkinEST08,AbrahamBN08}, in the spirit of the backbone reduction of~\cite{GuptaKR09} but with a different accounting.

\begin{theorem}[Online $\{1,2\}$-VC-SNDP]
\label{thm:k1-intro}
Online $\{1,2\}$-vertex-connectivity SNDP, where each demand pair requests connectivity or $2$-vertex-connectivity, admits a polynomial-time randomized $\widetilde O(\log^3 n)$-competitive algorithm.
\end{theorem}

To the best of our knowledge, \Cref{thm:k1-intro} is the first polylogarithmic-competitive online algorithm for any vertex-connectivity SNDP variant. See \Cref{thm:VCSNDPfull} for a detailed version of \Cref{thm:k1-intro}.

\paragraph{Organization of the paper.}
We survey the related work in \Cref{sec:related}.
\Cref{sec:overview} gives a technical overview of the shattering dichotomy, the Containment Lemma, and the algorithmic reductions.
\Cref{sec:prelim} fixes notation and states the online and offline primitives that we utilize. \Cref{sec:framework} develops the component cut feasibility framework, gives the reductions for shattering and non-shattering cuts, and presents the deterministic and randomized online algorithms for OVCA as well as the offline algorithm.
\Cref{sec:1-2-online} gives the online algorithm for $\{1,2\}$-VC-SNDP. 
In \Cref{sec:hardness} we discuss the hardness of OVCA.
\Cref{sec:conclusion} lists open problems.
Lastly, \Cref{sec:necessity} gives an example demonstrating the necessity of our shattering dichotomy.

\subsection{Related Work}
\label{sec:related}

\paragraph{Online survivable network design and Steiner primitives.}
For online edge-connectivity SNDP, Gupta, Krishnaswamy, and Ravi \cite{GuptaKR09} obtained an $O(k\log^3 n)$-competitive randomized algorithm.
Dehghani, Ehsani, Hajiaghayi, Liaghat, and Seddighin \cite{DehghaniEHLS18} gave a $(O(\log^2 n \log_{1+\varepsilon} k), 2+\varepsilon)$-bicriteria competitive algorithm: the solution achieves a connectivity of at least $r_i / (2+\varepsilon)$ for every requirement $r_i$, and its total cost is not more than $O(\log^2 n \log_{1+\varepsilon} k)$ times the cost of an optimal solution satisfying all connectivity requirements.
Online Steiner tree and forest have $O(\log n)$-competitive algorithms~\cite{ImaseW91,BermanC97}, with matching lower bounds.
Early polylogarithmic-competitive algorithms for online node-weighted Steiner tree and forest were given by Naor, Panigrahi, and Singh~\cite{NaorPS11} and Hajiaghayi, Liaghat, and Panigrahi~\cite{HajiaghayiLP13}, respectively.
For online node-weighted Steiner forest, Borst, Eli\'a\v{s}, and Venzin \cite{BorstEV25} obtained the near-optimal randomized ratio $O(\log k\cdot\log n)$.
Crucially for us, when the set of possible terminal pairs is fixed in advance they give a \emph{deterministic} $O(\log|T|\cdot\log N)$-competitive algorithm, where $N$ is the number of nodes and $T$ is the set of possible demands.
Online set cover and hitting set have $O(\log m\cdot\log n)$-competitive algorithms, which can be made deterministic when the set system is fixed in advance \cite{AlonAABN09}; the underlying online primal-dual framework~\cite{AlonAABN06,BuchbinderN09} also yields the fractional and randomized covering guarantees that we use to remove the dependence on~$k$ from the running time.

\paragraph{Online vertex connectivity.}
To the best of our knowledge, the only prior work on online vertex-connectivity network design is a side result in the work of Naor, Panigrahi, and Singh~\cite{NaorPS11}. They considered the edge-weighted single-source setting and gave a bicriteria algorithm. More precisely, for any $\epsilon>0$, connectivity requirement $\ell$, and fixed source $s$, their deterministic polynomial-time online algorithm maintains a subgraph in which every arrived online terminal $t_i$ has vertex connectivity at least $\ell/(2+\epsilon)$ to $s$. The cost is $O\!\left(\frac{\ell \log n}{\epsilon}\right)$-competitive with the optimum offline subgraph that guarantees $\ell$-vertex-connectivity from $s$ to every terminal.

\paragraph{Online edge-connectivity augmentation.}
Naor, Umboh, and Williamson~\cite{NaorUW22} gave a tight deterministic $O(\log n)$-competitive algorithm for online weighted tree augmentation, the $k=1$ case of edge-connectivity augmentation.
Closest to our work is that of Garg and Subramanian~\cite{GargS26}, who study the \emph{edge}-connectivity analogue: starting from a $k$-edge-connected graph, and receiving the demand pairs online, augment them so that demanded pairs become $(k{+}1)$-edge-connected.
They obtain a tight deterministic $O(\log n)$ ratio in the unweighted case via online tree augmentation and edge-weighted Steiner forest, and $O(\log^2 n)$ in the weighted case via online node-weighted Steiner forest on an auxiliary graph derived from the cactus representation of minimum edge cuts.
Although both \cite{GargS26} and our work ultimately invoke the same online node-weighted Steiner forest primitive of Borst, Eli\'a\v{s}, and Venzin~\cite{BorstEV25}, they use it in fundamentally different ways.
Their weighted auxiliary graph has paid link nodes, zero-cost cactus-vertex nodes, and edges encoding incidence and crossing links.
For general~$k$, minimum vertex cuts have no analogous cactus representation. Our auxiliary graph also shares paid candidate-edge nodes, but uses component nodes indexed by shattering cuts, with one demand generating multiple Steiner-forest requests. Its soundness requires controlling auxiliary paths that traverse nodes belonging to different cuts.
Their ratio is independent of~$k$, whereas our \Cref{thm:hardness-intro} rules out a $k$-independent $\polylog(n)$ competitive ratio for OVCA.
This shows a clear-cut separation between the edge-connectivity and vertex-connectivity settings.

Garg, Hommelsheim, and Lindermayr~\cite{GargHL26} subsequently improved the weighted case to an optimal deterministic polynomial-time $O(\log n)$-competitive algorithm.
Their techniques differ from the Steiner forest approach: they construct a laminar family of minimum cuts online and use primal-dual chain covering, rather than reducing to node-weighted Steiner forest.

Hathcock, Subramanian, Umboh, and Zlatin~\cite{HathcockSUZ26} also study online edge-connectivity augmentation in a different \emph{cut-arrival} model, where individual minimum edge cuts arrive as requests, using a reduction to geometric hitting set.

\paragraph{Offline vertex-connectivity network design and augmentation.}
Chuzhoy and Khanna \cite{ChuzhoyK09} gave an $O(k^3\log n)$-approximation for VC-SNDP.
The special case of $\{0,1,2\}$ requirements has a deterministic $2$-approximation \cite{Fleischer01VC012}.
Hardness is strong, both in~$n$ \cite{KortsarzKL04} and in~$k$.
Laekhanukit \cite{Laekhanukit14} proved a $k^{1/6-\varepsilon}$ bound, improved by Manurangsi \cite{Manurangsi19} to $\tilde\Omega(k^{1/5})$ and $\tilde\Omega(D^{1/2})$ in the number of demands.
These bounds concern \emph{general} VC-SNDP and do not directly apply to augmentation from a free $k$-connected backbone.

Augmentation-by-one itself is classical: Frank~\cite{Frank92} solved edge-connectivity augmentation,
Jord\'an~\cite{Jordan95} gave additive approximation,
Jackson and Jord\'an~\cite{JacksonJ05}
gave an algorithm to solve the
undirected vertex connectivity augmentation problem exactly for any fixed $k$, and finally V\'egh~\cite{Vegh11} gave an exact polynomial-time algorithm for the same; see also~\cite{VeghB08} for the directed case.
Thus the complete, \emph{uniform-cost complete-graph} version of augmenting \emph{all pairs} from $k$ to $k{+}1$ is polynomial-time solvable.
This tractability is however quite brittle:
Kortsarz, Krauthgamer, and Lee~\cite{KortsarzKL04} proved APX-hardness for all-pairs vertex connectivity augmentation in the case where costs are uniform but the candidate graph is not complete,
as well as in the case where costs are in $\{1,2\}$ and the candidate graph is complete.
See also~\cite{MaKWWSW10} for hardness of global augmentation.
Problems involving selected-pair connectivity augmentation are at least as hard as Steiner forest, thus also APX-hard~\cite{bern1989steiner}.
Approximation algorithms are known for various weighted and subset variants~\cite{KortsarzN08,Nutov24,Nutov11SubsetAug}.
For offline augmentation of all vertex pairs from $k$ to $k+1$, Nutov~\cite{nutov2018improved} gives an $O(1+\log(n/(n-k)))$-approximation.

For unit-cost links and global vertex-connectivity requirements, Carmesin and Ramanujan~\cite{CarmesinR26} give an FPT algorithm, parameterized by the number of added links, for target connectivity at most~$4$.
Korhonen and Thorup~\cite{KorhonenT26} give an exact $2^{O(b\log(b+\ell))}n^{O(1)}$-time algorithm for augmenting an arbitrary graph to global $\ell$-vertex-connectivity using at most $b$ candidate links.
These results concern offline global connectivity, whereas we study weighted augmentation for selected demand pairs.

Reductions from connectivity augmentation to Steiner problems also appear in offline algorithms.
Basavaraju, Fomin, Golovach, Misra, Ramanujan, and Saurabh~\cite{BasavarajuFGMRS14} reduce edge-connectivity augmentation to node-weighted Steiner tree, while Nutov~\cite{NutovESA24} reduces rooted vertex-connectivity augmentation to node-weighted group Steiner tree.
These constructions exploit the tree/cactus structure of minimum edge cuts or the intersecting family of tight bisets associated with a common root; neither structure is available for arbitrary-pair higher-order vertex-connectivity augmentation.
Thus the reduction to a Steiner problem is not new as a general paradigm; our contribution is a shared online node-weighted Steiner forest construction for shattering minimum vertex cuts.
Its soundness is not automatic: an auxiliary path connecting two components of $F-C$ may pass through component-nodes belonging to other cuts, and the corresponding route in the original graph may pass through $C$ itself.
The Containment Lemma makes such paths safe for shattering cuts (\Cref{lemma:FeasibleNWSFtoVCA}); without this restriction, the construction can certify an infeasible augmentation, as our counterexample in \Cref{sec:necessity} shows.

\paragraph{Directed Steiner network and forest.}
Directed Steiner Network (equivalently Directed Steiner Forest) is a well-known hard problem.
Dodis and Khanna~\cite{DodisK99} proved hardness ruling out fixed-polylogarithmic approximation unless $\mathrm{NP}$ has quasi-polynomial-time algorithms, and Hoppenworth, Long, Mahabadi, and Tarnawski~\cite{HLMT27} showed a similar unconditional hardness for online algorithms.
The best polynomial-time (offline) ratios are $O(D^{1/2+\varepsilon})$~\cite{ChekuriEGS11,FeldmanKN12} and $O(n^{2/3+\varepsilon})$~\cite{BermanBMRY13} for general costs, and $\widetilde O(n^{4/7+\varepsilon})$ for uniform costs~\cite{AbboudB18}. Online, the best known competitive ratios are $\widetilde O(D^{1/2+\varepsilon})$ for general costs~\cite{ChakrabartyEKP18} and $\widetilde O(n^{3/5+\varepsilon})$ for uniform costs~\cite{GrigorescuLQ21,DBLP:conf/soda/BodwinL25}.
The reduction of \Cref{thm:ReductionDSNtoVCA} places OVCA above this problem.

\section{Technical Overview}
\label{sec:overview}

This section gives a high-level, self-contained explanation of why online vertex connectivity augmentation is difficult, how we overcome the difficulties, and our main technical contributions.

\subsection{Repairing Component Cuts}
\label{ssec:ov-repair}

Fix a demand pair $(s,t)$ and for simplicity, suppose that $s$ and $t$ are non-adjacent in~$F$.
By Menger's theorem, making $(s,t)$ $(k{+}1)$-vertex-connected in $F\cup X$ means ensuring that {no} set of $k$ vertices separates them.
Since $F$ is already $k$-connected, the only sets of $k$ vertices that could still separate $s$ from $t$ are the minimum cuts of~$F$.
Thus the task is:

\begin{quote}
\emph{For every $k$-vertex cut $C$ of $F$ that separates $s$ from $t$, the purchased edges $X$ must reconnect $s$ to $t$ in $(F\cup X)-C$.}
\end{quote}

It is convenient to look at one cut $C$ through its \emph{component quotient}.
Contract each connected component of $F-C$ to a single node.
An edge $(a,b)$ with $a,b\notin C$ (we say $(a,b)$ avoids $C$) becomes an edge between the nodes of the two components containing $a$ and~$b$. Edges of $F$ disappear inside the contraction.
Then $s$ and $t$ are reconnected in $(F+X)-C$ if and only if their component-nodes are connected using only avoiding edges of $X$.
Equivalently, \emph{every} bipartition of the components that puts $s$'s component on one side and $t$'s on the other must be crossed (or \emph{covered}) by some avoiding edge of~$X$.
We call such a bipartition $(L,R)$ of $V - C$, together with its vertex cut $C$, a \emph{component cut}, denoted $(L,C,R)$. Then the repair problem is exactly to cover every component cut with an avoiding edge.

This already suggests a naive algorithm: each component cut is a covering constraint, so feed all of them to an online hitting-set (set-cover) subroutine.
The issue is the large number of constraints.
A cut $C$ with $\ell$ components generates one constraint for every way of splitting the other $\ell-2$ components across the two sides, i.e.\ $2^{\ell-2}$ component cuts,
and $\ell$ can be as large as $\Theta(n)$.
So the naive hitting-set encoding produces up to $2^{\Theta(n)}$ constraints from one cut,
which is prohibitively large
even when we use hitting-set subroutines where one pays the logarithm of this quantity. This is exactly the main difficulty when dealing with vertex connectivity. In the edge connectivity setting, in contrast, deleting a minimum edge cut only leaves two components.

\subsection{The Shattering Dichotomy and the Containment Lemma}
\label{ssec:ov-shattering}

Our key idea is to classify cuts by exactly this quantity -- the number of components they create -- and to handle the two regimes by completely different online primitives.

\begin{itemize}
\item A minimum vertex cut $C$ of $F$ is \textbf{non-shattering} if $F-C$ has at most $k$ components.
Then each such cut spawns only $2^{\ell-2}\le 2^{k-2}$ component cuts.
This is exponential in~$k$ but independent of~$n$.
We feed these constraints to online hitting set.

\item A minimum vertex cut $C$ of $F$ is \textbf{shattering} if $F-C$ has at least $k{+}1$ components.
Now the hitting-set encoding is too heavy, and we must proceed differently.
We route all shattering cuts to a single, shared online \emph{node-weighted Steiner forest} instance.
We are then able to post only a single constraint per shattering vertex cut.
\end{itemize}

\paragraph{The auxiliary Steiner forest.}
Build a bipartite node-weighted graph~$H$.
On one side, put a zero-weight node for every component of $F-C$, over all shattering vertex cuts~$C$.
On the other side, put a node of weight $c(e)$ for every candidate edge $e\in U$. For each candidate edge $e$ and each shattering vertex cut $C$, if $e$ connects two different components of $F-C$, add edges from the $e$-node to these two component-nodes.
When a demand $(s,t)$ arrives, for each shattering cut $C$ separating them we post just one Steiner-forest demand: connect the component of $s$ to the component of~$t$.
Purchasing the candidate edges selected by the Steiner-forest algorithm is our repair for all shattering cuts at once.

To show that cut $C$ is repaired for demand $(s,t)$,
we look at the path
connecting $s$'s $C$-component to $t$'s $C$-component
in the Steiner forest solution in $H$.
The crucial subtlety is that
this path is free to wander through component-nodes of \emph{other} shattering cuts $D\ne C$.
A priori, such a wandering path could route through a component of $F-D$ that actually {contains} a vertex of~$C$ -- in which case deleting~$C$ may sever the route, and the repair would fail.
We argue that this does not happen,
starting from
the following rigidity property of shattering cuts,
which is a dual form of a lemma of Jord\'an~\cite{Jordan95}.

\begin{quote}
\textbf{Containment Lemma (\Cref{lemma:ContainmentLemma}).}
\emph{If $C$ is a shattering minimum vertex cut and $D$ is any minimum vertex cut, then $C\subseteq D\cup \gamma$ for a single component $\gamma$ of $F-D$.}
\end{quote}

The proof is simple. Each vertex of a minimum cut $C$ has a neighbor in every $(F-C)$-component (otherwise removing the other $k-1$ vertices would already disconnect that component from the others, violating $k$-connectivity).
So between any two vertices $x,y\in C$ we can route one internally vertex-disjoint path through each of the (at least) $k{+}1$ components of $F-C$.
If $C$ had vertices on two different sides of~$D$ -- say $x$ and $y$ -- then each of those $k{+}1$ paths would have to cross~$D$, using $k{+}1$ distinct vertices of~$D$; but $|D|=k$, a contradiction.

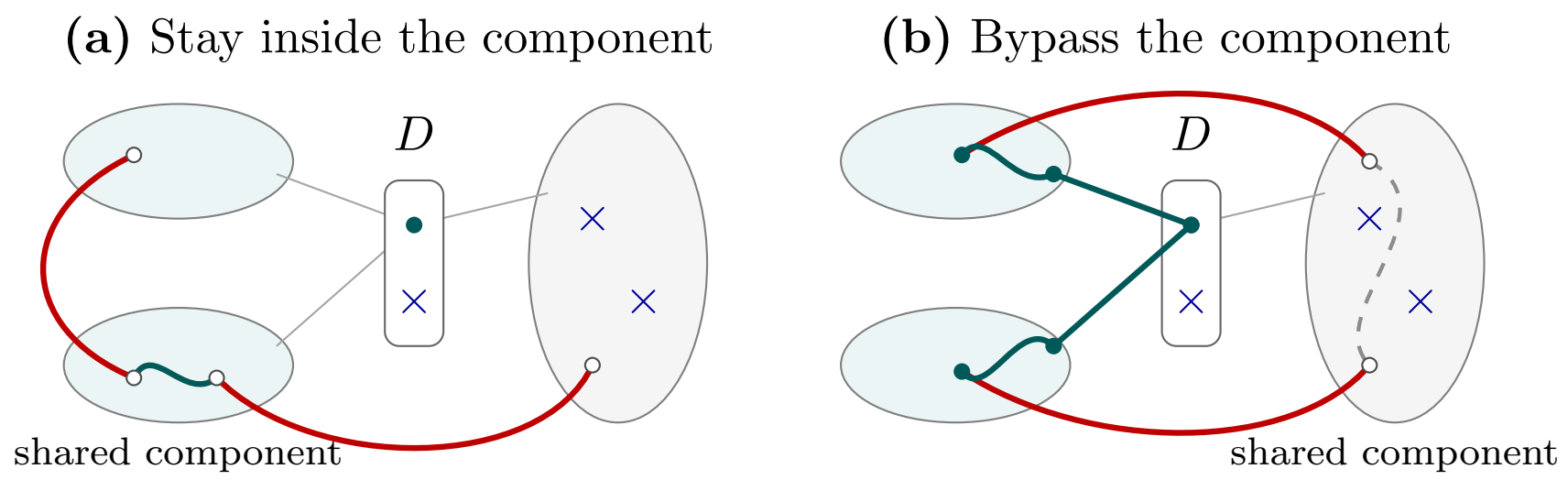
\begin{figure}[t]
\centering
\begin{tikzpicture}[
    scale=0.92, font=\small,
    component/.style={draw=black!50, fill=teal!8},
    affected/.style={draw=black!50, fill=black!4},
    cut/.style={draw=black!60, rounded corners=3pt, fill=white},
    vertex/.style={circle, draw=black!70, fill=white, inner sep=0pt, minimum size=3pt},
    reached/.style={vertex, draw=teal!70!black, fill=teal!70!black},
    deleted/.style={blue!60!black, inner sep=0pt},
    purchased/.style={red!75!black, very thick},
    surviving/.style={teal!70!black, very thick},
    label text/.style={font=\scriptsize}
]
\foreach \offset in {0, 6.1} {
\begin{scope}[xshift=\offset cm]
    \draw[component] (-1.85,0.8) ellipse (0.9 and 0.45);
    \draw[component] (-1.85,-0.8) ellipse (0.9 and 0.45);
    \draw[affected] (1.6,0) ellipse (0.7 and 1.25);
    \draw[black!35] (-1.08,0.7) -- (0,0.3) -- (-1.08,-0.65);
    \draw[black!35] (0,0.3) -- (1.05,0.55);
    \draw[cut] (-0.23,-0.65) rectangle (0.23,0.65);
    \node at (0,1.02) {$D$};
    \node[reached] at (0,0.3) {};
    \node[deleted] at (0,-0.3) {$\times$};
    \node[deleted] at (1.4,0.35) {$\times$};
    \node[deleted] at (1.8,-0.3) {$\times$};
\end{scope}
}
\node[align=center, text width=5.4cm] at (-0.2,1.75)
    {\textbf{(a)} Stay inside the component};
\node[label text] at (-1.85,-1.5) {shared component};
\draw[purchased] (-2.2,0.85) .. controls (-3.15,0.4) and (-3.15,-0.5) .. (-2.2,-0.9);
\draw[purchased] (-1.55,-0.9) .. controls (-0.8,-1.65) and (1.0,-1.65) .. (1.4,-0.8);
\draw[surviving] (-2.2,-0.9) .. controls (-2.05,-0.6) and (-1.8,-1.1) .. (-1.55,-0.9);
\foreach \point in {(-2.2,0.85),(-2.2,-0.9),(-1.55,-0.9),(1.4,-0.8)}
    \node[vertex] at \point {};
\begin{scope}[xshift=6.1cm]
    \node[align=center, text width=5.4cm] at (-0.2,1.75)
        {\textbf{(b)} Bypass the component};
    \node[label text] at (1.6,-1.5) {shared component};
    \draw[purchased] (-1.8,0.85) .. controls (-0.8,1.5) and (0.8,1.5) .. (1.4,0.8);
    \draw[purchased] (-1.8,-0.85) .. controls (-0.8,-1.5) and (0.8,-1.5) .. (1.4,-0.8);
    \draw[black!45, thick, dashed] (1.4,0.8) .. controls (2.1,0.4) and (1.0,-0.4) .. (1.4,-0.8);
    \draw[surviving] (-1.8,0.85) .. controls (-1.6,1.1) and (-1.4,0.55) .. (-1.08,0.7)
        -- (0,0.3) -- (-1.08,-0.65)
        .. controls (-1.35,-0.4) and (-1.6,-1.1) .. (-1.8,-0.85);
    \foreach \point in {(-1.8,0.85),(-1.08,0.7),(0,0.3),(-1.08,-0.65),(-1.8,-0.85)}
        \node[reached] at \point {};
    \foreach \point in {(1.4,0.8),(1.4,-0.8)}
        \node[vertex] at \point {};
\end{scope}
\end{tikzpicture}
\caption{An idealized illustration of why containment makes reachability propagate correctly. Red curves are purchased edges, teal curves are paths in $F-C$, and crosses are deleted vertices of $C$. If the shared component of $F-D$ avoids $C$, connect inside it (a); otherwise, use the other endpoints and bypass it through the surviving part of $D$ (b). (Only selected vertices and components are shown.)}
\label{fig:containment-overview}
\end{figure}

The Containment Lemma is the main structural reason why the shared Steiner forest reduction is sound; see \Cref{fig:containment-overview}.
When a Steiner path for cut $C$ passes through a component-node of another cut $D$, the lemma confines the dangerous vertices $C\setminus D$ to a single component~$\gamma$ of $F-D$; the union of the \emph{other} components of $F-D$ with $D\setminus C$ then forms a connected region that entirely avoids~$C$, and reachability propagates through it.
Iterating along the Steiner path turns connectivity in the auxiliary graph~$H$ into a true $s$--$t$ path in $(F\cup X)-C$.
We carry out this induction in \Cref{lemma:FeasibleNWSFtoVCA}.

Therefore,
a feasible solution of the single auxiliary instance~$H$ repairs \emph{every} shattering vertex cut (and its corresponding component cuts) simultaneously.
The shattering story is completed by the fact
that the number of such cuts can be bounded by $O(n)$.
Thanks to this, the auxiliary instance~$H$ is polynomial-size.

\paragraph{Tightness, and why a weaker hypothesis fails.}
One might hope the same reduction works for any cut,
but it does not.
The counting argument in the proof of the Containment Lemma needs $k{+}1$ components to overload the $k$ vertices of~$D$,
and this is in fact indispensable,
as already for $k$-component cuts the Lemma fails, and so does the correctness of the reduction.
In \Cref{sec:necessity} we show a $4$-connected graph with a $4$-component cut $C$
(not shattering, since $4<k{+}1=5$)
and a second cut~$D$ for which $C$ straddles {both} components of $F-D$.
We also give a legal path in the auxiliary graph such that purchasing the corresponding original candidate edges does \emph{not} reconnect $s$ and~$t$ after deleting~$C$.
This demonstrates that non-shattering cuts must be dealt with differently.

\subsection{Covering Non-Shattering Component Cuts}
\label{ssec:ov-component-cuts}

The non-shattering side is governed by the component cuts $(L,C,R)$ introduced in \Cref{ssec:ov-repair}: a vertex cut~$C$ and a bipartition $(L,R)$ of $V-C$ with $s\in L$, $t\in R$ respecting the components of $F-C$.
It is a natural consequence of Menger's theorem that an augmentation $X$ makes $(s,t)$ $(k{+}1)$-connected if and only if $X$ has an avoiding edge crossing {every} component cut for $(s,t)$
(see \Cref{lem:component-cut-exact}).
Moreover, it turns out that the number of all component cuts arising from non-shattering vertex cuts is $4^k\cdot \poly(n)$\footnote{In fact, a slightly more careful analysis gives $2^{k}\cdot\poly(n)$.} (\Cref{lemma:NonShatteringCutsNumber}).
The exact way we deal with these component cuts depends on the algorithmic setting:

\begin{enumerate}
\item \textbf{Deterministic online in $\poly(2^k,n)$ time.}
Enumerate all minimum cuts, which are $O(2^k\,n^2)$ many~\cite{Kanevsky90}, split them into shattering and non-shattering, feed the (at most $2^{k-2}$ per cut) non-shattering component cuts to deterministic online hitting set~\cite{AlonAABN09}, and route the shattering cuts through deterministic online node-weighted Steiner forest~\cite{BorstEV25} as outlined above.
This yields a competitive ratio of $O((k+\log n)\log n)$.

\item \textbf{Randomized online in polynomial time.}
To avoid the $2^k$ enumeration, which is only polynomial-time when $k \le O(\log n)$, we maintain the component cut constraints \emph{fractionally}.
We use an oracle-efficient online covering framework (\cite{AlonAABN06,BuchbinderN09,GrigorescuLQ21} -- see \Cref{thm:online-covering}) to maintain a good fractional solution while encountering only polynomially many violated constraints. The required separation oracle is implemented using a single max-flow computation.
We then round online by independent {threshold layers} at scale $\Theta(k+\log n)$, never enumerating the exponential family.
This pays $O((k+\log n)\log n)$ in expectation and yields a randomized Las~Vegas algorithm that works against an oblivious adversary.

\item \textbf{Offline.}
We proceed similarly as the randomized polynomial-time online algorithm, but replace the online fractional covering LP solver with the ellipsoid method, and also replace the online node-weighted Steiner forest subroutine with its $O(\log n)$-approximate offline algorithm \cite{KleinRavi95}. These together shave a $\log n$ factor in the approximation factor.

\end{enumerate}

The two regimes are thus not interchangeable: hitting set / LP rounding cannot handle shattering cuts (too many components, and rounding fails to achieve a good approximation ratio), while the Steiner forest reduction is provably incorrect on non-shattering cuts.
The dichotomy is what lets each primitive run exactly where it is sound.

\subsection{$\{1,2\}$-Vertex-Connectivity SNDP}
\label{ssec:ov-k1}

When we assume there is a free spanning tree, the online $\{1,2\}$-VC-SNDP problem becomes an OVCA problem with $k=1$. To remove the free-tree assumption, we borrow ideas from the online EC-SNDP algorithm by Gupta, Krishnaswamy, and Ravi~\cite{GuptaKR09}, which exploits a probabilistic low-stretch spanning tree sampler \cite{AbrahamBN08}. The benefit of considering a low-stretch spanning tree $T$ is that, whenever we plan to purchase an edge $e$, it is affordable to simultaneously purchase the whole fundamental cycle of $e$. This makes the tree $T$ effectively free without buying the whole tree in advance (see \Cref{lemma:EffectivelyFree}).

The algorithm is then straightforward. When a demand pair $(s,t)$ arrives, buy the $s$-$t$ tree path. If $(s,t)$ has connectivity requirement $2$, feed it to the OVCA algorithm assuming $T$ is free, where each non-tree edge $e$ is charged the inflated cost $c_T(e)$ of its whole fundamental cycle (see the proof of \Cref{thm:VCSNDPfull} for details). When the OVCA algorithm purchases an edge, we actually buy that fundamental cycle instead. This gives a competitive ratio of $\widetilde{O}(\log^{3}n)$: an $\widetilde{O}(\log n)$ factor from the low-stretch spanning tree, and the remaining $O(\log^{2}n)$ factor from the online OVCA algorithm.

\subsection{Hardness}
\label{ssec:ov-hardness}

Could one hope for a $k$-independent $\polylog(n)$ ratio?
The answer is negative. Somewhat surprisingly,
it turns out~\cite{LandoN09} that the (undirected) OVCA problem already contains \emph{directed} reachability.
More precisely,
there is an exact, cost- and demand-preserving reduction from Directed Steiner Network with $k=n/2$, so the known hardness of that problem~\cite{DodisK99,HLMT27} rules out $k$-independent polylogarithmic approximation.
The reduction also calibrates the positive side: any $O(k^{\alpha}\polylog n)$ ratio with $\alpha<2/3$ would improve the long-standing frontier for offline Directed Steiner Forest \cite{BermanBMRY13},
and an online algorithm with $\alpha < 1$ would give the first online algorithm with truly sublinear competitive ratio for Directed Steiner Forest with general edge weights.

\section{Preliminaries}
\label{sec:prelim}

In \Cref{ssec:prelim-graphs}, we introduce basic graph-theoretic notation. \Cref{ssec:ProblemPrelim} introduces notation related to the input OVCA instance. \Cref{ssec:prelim-cuts} introduces notation and several useful structural properties about vertex cuts. In \Cref{ssec:prelim-blackboxes}, we state several online and offline subroutines from prior work that will be used by our algorithms as black boxes.

\subsection{Basic Graph Notation}
\label{ssec:prelim-graphs}

\paragraph{Graphs.}
All graphs are undirected and simple.
For a graph $G$, we use $V(G)$ and $E(G)$ to denote the vertex and edge sets of $G$.
For a subset of vertices $C\subseteq V(G)$, $G-C$ is the subgraph induced by $V(G)\setminus C$. To avoid clutter, we define $G-v:=G-\{v\}$ for each vertex $v$, and $G-e:=G-\{e\}$ for each edge $e$. We say an edge $e=(u,v)$ \emph{avoids} a vertex set $C$ if $u,v\notin C$.

We use $\Gamma(G)$ to denote the set of connected components of $G$, and for each vertex $v$ in $G$, let $\gamma(v,G)\in \Gamma(G)$ denote the connected component containing $v$. 

\paragraph{Vertex Cuts.} A \emph{vertex cut} of $G$ is a vertex set $C\subseteq V(G)$ such that $G-C$ is disconnected, and $C$ is a minimum vertex cut of $G$ if its size $|C|$ is the smallest among all vertex cuts. We say $C$ is a $k$-vertex cut when it has size exactly $k$.

\paragraph{Vertex Connectivity.}
For $s,t\in V(G)$, we define the vertex connectivity $\kappa_G(s,t)$ as the maximum number of internally vertex-disjoint $s$-$t$ paths.
By Menger's theorem~\cite{Diestel17}, if $s$ and $t$ are non-adjacent, then $\kappa_G(s,t)$ equals the minimum size of a vertex cut $C\subseteq V(G)\setminus\{s,t\}$ whose removal separates $s$ from $t$.
When $s$ and $t$ are adjacent, the edge $(s,t)$ is a path with no internal vertex, and
\begin{equation*}
\kappa_G(s,t)=\kappa_{G-(s,t)}(s,t)+1 .
\end{equation*}
A graph is \emph{$k$-(vertex-)connected} if $\kappa_G(s,t)\ge k$ for all vertex pairs $s,t\in V(G)$.

\begin{remark}
We note that vertex connectivity between $s$ and $t$ is sometimes defined as the minimum size of a vertex cut that separates $s$ and $t$, even when $s$ and $t$ are adjacent (in which case it becomes essentially $+\infty$). We use the disjoint-paths definition as it is more standard in network design literature (see e.g.~\cite{ChuzhoyK09}).
However, our results would also hold under the alternative definition; in fact, it would simplify the setting as adjacent demands would become vacuous.
\end{remark}

\subsection{The OVCA Instance}
\label{ssec:ProblemPrelim}

We use the notation of \Cref{def:problem} throughout. The input graph is $G=(V,E)$ with edge cost function $c$. $F$ is the free graph, $U=E\setminus E(F)$ is the set of candidate edges, and $X\subseteq U$ denotes the set of purchased edges maintained by an algorithm (we also call $X$ an \emph{augmentation}). We use $\OPT_{\VCA}$ to denote the \emph{offline} optimum of this OVCA instance.

A demand pair $d=(s,t)\in\binom{V}{2}$ is adjacent if $(s,t)\in E(F)$, and non-adjacent otherwise.
To avoid clutter, we introduce the following uniform notation for adjacent and non-adjacent demand pairs.
Define
\begin{equation}
\label{eq:Fd}
F_d:=\begin{cases}F,& (s,t)\notin E(F),\\ F-(s,t),& (s,t)\in E(F),\end{cases}
\qquad
r_d:=\begin{cases}k+1,& (s,t)\notin E(F),\\ k,& (s,t)\in E(F).\end{cases}
\end{equation}
The above notation reduces an adjacent pair in $F$ to a non-adjacent pair in $F_{d}$ in the following sense: augmenting the vertex connectivity of the demand pair $(s,t)$ to $k+1$ in $F\cup X$ is equivalent to augmenting the connectivity of $(s,t)$ to $r_{d}$ in $F_{d}\cup X$. This is formally stated in the following \Cref{ob:AdjacentReduction}.

\begin{observation}
\label{ob:MinCutSize}
For each adjacent demand pair $d$, $F_{d}$ is $(k-1)$-vertex-connected.
\end{observation}

\begin{observation}
\label{ob:AdjacentReduction}
Let $X\subseteq U$ be any augmentation. For each demand pair $d = (s,t)$, $\kappa_{F\cup X}(s,t)\geq k+1$ if and only if $\kappa_{F_{d}\cup X}(s,t)\geq r_{d}$.
\end{observation}

Lastly, we make the following assumptions on the input instance without loss of generality. 
\begin{assumption}
\label{assume:Input}
We assume the following.
\begin{enumerate}
\item\label{item:InputAssume1} The graph $G$ is a complete graph, so every demand pair $d = (s,t)\in \Dcal$ can be made $(k+1)$-vertex-connected using the candidate edges $U$.
\item\label{item:InputAssume2} Every demand pair $d=(s,t)\in \Dcal$ has $\kappa_{F}(s,t) = k$.
\end{enumerate}
\end{assumption}

We now argue that the above assumptions are without loss of generality. For \Cref{item:InputAssume1}, we can add any missing edge to $U$ with sufficiently large cost $+\infty$, so that $G = F + U$ is a complete graph. For \Cref{item:InputAssume2}, we can simply filter out all demand pairs $(s,t)$ with $\kappa_{F}(s,t)>k$ (this can be checked in polynomial time).

\subsection{Structural Properties of Vertex Cuts}
\label{ssec:prelim-cuts}

\begin{lemma}[\cite{Kanevsky90}]
\label{fact:num-cuts}
In a $k$-vertex-connected graph on $n$ vertices, the number of $k$-vertex cuts is at most $O(2^{k}\cdot n^{2})$, and there is a deterministic algorithm that enumerates these vertex cuts in $O(2^{k}\cdot n^{3})$ time.
\end{lemma}

In a graph $G$, a \emph{$k$-shredder} $C$ is a $k$-vertex cut such that there are three or more connected components in $G-C$, i.e. $|\Gamma(G-C)|\geq 3$.

\begin{lemma}[\cite{CheriyanT99,LibermanN07}]
\label{fact:shredders}
In a $k$-vertex-connected graph, the number of $k$-shredders is at most $O(n)$, and there is a deterministic algorithm that lists them explicitly in polynomial time.
\end{lemma}

\subsection{Online and Offline Subroutines}
\label{ssec:prelim-blackboxes}

Next, we state several online and offline algorithms that will be used as subroutines in our algorithms. In each theorem statement in this subsection, we use $\OPT$ to denote the offline optimum of the online instance.

\paragraph{Online Node-Weighted Steiner Forest.}

In the online node-weighted Steiner forest (ONWSF) problem, the input consists of a graph $H$ with node weights $w:V(H)\to \mathbb{R}_{\geq 0}$, and an online sequence $\mathcal{T}$ of terminal pairs $(x_{1},y_{1}),(x_{2},y_{2}),\ldots$ in $\binom{V(H)}{2}$. We need to maintain an incremental (i.e., nodes cannot be removed from $S$) node set $S\subseteq V(H)$ such that, at any time $i$, the arrived pair of terminals $(x_{i},y_{i})$ must be connected in $H[S]$. The goal is to minimize the total weight of nodes in $S$, denoted by $w(S)$.

\begin{theorem}[Deterministic ONWSF, {\cite[Theorem~2]{BorstEV25}}]
\label{thm:onwsf}
Given a node-weighted graph $H$ with $N$ vertices and an online sequence of terminal pairs, there is a deterministic algorithm that maintains an ONWSF solution $S$ with competitive ratio $O(\log^{2}N)$, i.e. $w(S)\leq O(\log^{2}N)\cdot \OPT$ at the end.
The algorithm runs in $\poly(N)$ time.
\end{theorem}

In \Cref{ssec:shattering-reduction}, we will apply \Cref{thm:onwsf} on a graph $H$ with $N=O(n^{2})$ vertices, so this subroutine will take $\poly(n)$ time.

\paragraph{Online Set Cover and Hitting Set.} In the online set cover problem, the input consists of
\begin{itemize}
\item a ground set $X$ of $N$ elements and a family $\mathcal{A}\subseteq 2^{X}$ of $M$ sets with costs $c:\mathcal{A}\to \mathbb{R}_{\geq 0}$ (i.e., both $X$ and $\mathcal{A}$ are known in advance), and
\item an online sequence of elements.
\end{itemize}
We need to maintain an incremental subfamily $\mathcal{B}$ of $\mathcal{A}$ such that at any time, each arrived element must lie in some set in $\mathcal{B}$. The goal is to minimize the total cost of sets in $\mathcal{B}$.

\begin{theorem}[Deterministic online set cover, {\cite{AlonAABN09}}]
\label{thm:online-set-cover} Given the aforementioned ground set $X$, set family $\mathcal{A}$ and online sequence of elements, there is a deterministic algorithm that maintains an online set cover solution $\mathcal{B}$ with competitive ratio $O(\log N\cdot\log M)$, i.e., $c(\mathcal{B})\leq O(\log N\cdot\log M)\cdot\OPT$ at the end. The algorithm runs in $\poly(N,M)$ time.
\end{theorem}

Our algorithms actually use the hitting-set form of online set cover, defined as follows. In the online hitting set problem, the input consists of
\begin{itemize}
\item a ground set $U$ of $N$ elements with cost $c:U\to \mathbb{R}_{\geq 0}$, and a family $\mathcal{R}\subseteq 2^{U}$ of $M$ sets, and 
\item an online sequence of sets in $\mathcal{R}$.
\end{itemize}
We need to maintain an incremental subset $S$ of $U$ such that at any time, each arrived set $R$ is hit by some element in $S$, i.e. $S\cap R\neq\emptyset$. The goal is to minimize the total cost of elements in $S$.

\begin{corollary}[Deterministic online hitting set]
\label{cor:online-hitting-set}
Given the aforementioned ground set $U$, set family $\mathcal{R}$ and sequence of sets, there is a \emph{deterministic} algorithm that maintains an online hitting set solution $S$ with competitive ratio $O(\log N\cdot \log M)$, i.e., $c(S)\leq O(\log N\cdot\log M)\cdot \OPT$ at the end. The algorithm runs in $\poly(N,M)$ time.
\end{corollary}
\begin{proof}
Observe that it suffices to run \Cref{thm:online-set-cover} on the following online-set-cover instance with elements $X$ and set family $\mathcal{A}$. Let $X:=\mathcal{R}$ and $\mathcal{A}:=\{\mathcal{R}^{-1}(e)\mid e\in U\}$ where $\mathcal{R}^{-1}(e) = \{R\in\mathcal{R}\mid e\in R\}\subseteq X$ is the collection of sets in $\mathcal{R}$ containing $e$.
\end{proof}

\paragraph{Online Fractional Covering with Separation Oracles.}
Next, we state a polynomial-time online algorithm for solving online fractional covering LP, with access to a separation oracle.

Consider the following covering LP.

\begin{equation}
\label{eq:coveringlp}
\begin{aligned}
\min\qquad \mathbf{c}^{\top}\mathbf{x}&\\
\text{s.t.}\qquad 
\mathbf{a}^{\top}_{R}\mathbf{x}&\ge1
&& \forall R\in\mathcal{R}\\
\mathbf{x} &\ge \mathbf{0}
\end{aligned}
\end{equation}
Let $N$ be the number of variables. The input is as follows.
\begin{itemize}
\item The cost vector $\mathbf{c}$ is positive, which is given at the very beginning.\footnote{We purchase all zero-cost candidate edges at the outset and include them in the free baseline; hence every remaining LP variable has strictly positive cost.}
\item The covering constraints, indexed by $R\in\mathcal{R}$, arrive one by one in an online fashion, and all coefficient vectors $\abf^{\top}_{R}$ are non-negative. These constraints are not given explicitly and can only be accessed through a \emph{separation oracle}. At any time, given a candidate solution $\mathbf{x}$, the oracle either certifies that $\mathbf{x}$ satisfies all constraints that have arrived so far or returns a violated constraint.
\end{itemize}
We need to maintain an incremental LP solution $\xbf$ (namely, each variable is monotonically increasing over time) such that, at any time, all arrived constraints are satisfied. Naturally, the goal is to minimize the objective value $\cbf^{\top}\xbf$.

\begin{theorem}[Online fractional covering with separation oracles, {\cite[Theorem~4.1]{GrigorescuLQ21}}]
\label{thm:online-covering}
Given the aforementioned online LP instance, there is a deterministic online algorithm that maintains an incremental LP solution $\xbf$ that is feasible at any time with competitive ratio $O(\log N)$, i.e., $\cbf^{\top}\xbf\leq O(\log N)\cdot\OPT$ at the end. The algorithm queries the separation oracle 
\[
\poly\left(N,\log \OPT,\log\left(\frac{a_{\max}}{c_{\min}}\right)\right)
\]
times, where $c_{\min}$ is the minimum entry of the cost vector and $a_{\max}$ is the maximum coefficient among all coefficient vectors.
\end{theorem}

We note that the number of constraints does not affect either the competitive ratio or the running time guarantee in \Cref{thm:online-covering}. In our applications, we apply \Cref{thm:online-covering} to an LP with exponentially many constraints, $N=\poly(n)$ variables, and a separation oracle with $\poly(n)$ query time. Therefore, the algorithm achieves an $O(\log n)$ competitive ratio and makes only $\poly(n)$ calls to the separation oracle (and thus $\poly(n)$ running time).

\paragraph{Offline Node-Weighted Steiner Forest.} Lastly, we state an offline approximation algorithm for the node-weighted Steiner forest problem in \Cref{thm:klein-ravi}. In fact, \Cref{thm:klein-ravi} is a special case of the offline node-weighted network design algorithm by Klein and Ravi \cite{KleinRavi95}.

\begin{theorem}[Offline node-weighted Steiner forest, {\cite[Theorem~5.1]{KleinRavi95}}]
\label{thm:klein-ravi}
Given a node-weighted graph $H$ with $N$ vertices and a collection $\Tcal$ of terminal pairs, there is a deterministic algorithm that computes a set $S\subseteq V(H)$ such that every terminal pair is connected in $H[S]$. The algorithm is $O(\log N)$-approximate, i.e. the cost of $S$ is at most $O(\log N)\cdot \OPT$. The algorithm runs in polynomial time.
\end{theorem}

\section{Online Vertex Connectivity Augmentation Algorithms}
\label{sec:framework}

In this section, we will present two OVCA algorithms in \Cref{thm:deterministic} and \Cref{thm:randomizedOVCA}.

\begin{theorem}[Deterministic OVCA]
\label{thm:deterministic}
There is a deterministic online algorithm for OVCA with competitive ratio $O((k+\log n)\log n)$ and running time $\poly(2^k,n)$.
In particular, when $k=O(\log n)$, it runs in polynomial time and is $O(\log^2 n)$-competitive.
\end{theorem}

\begin{theorem}[Randomized OVCA]
\label{thm:randomizedOVCA}
There is a Las Vegas randomized online algorithm for OVCA with expected competitive ratio $O((k+\log n)\log n)$ and expected running time $\poly(n)$.
\end{theorem}

In \Cref{ssec:component-cuts}, we define the component cuts of demand pairs, and establish a feasibility criterion requiring all component cuts to be covered. Meanwhile, we classify these component cuts as either shattering or non-shattering. In \Cref{ssec:nonshattering}, we show how to cover all non-shattering component cuts by solving an online hitting set instance. In \Cref{sect:shattering}, we show how to cover all shattering component cuts by solving an online node-weighted Steiner forest instance.
In \Cref{sect:DeterministicOVCA}, we prove \Cref{thm:deterministic} using subroutines in \Cref{ssec:nonshattering} and \Cref{sect:shattering}.

To achieve polynomial running time, in \Cref{sect:RandomizedOVCA}, we replace the slower online hitting set subroutine with an online covering LP solver and an online rounding algorithm. This gives \Cref{thm:randomizedOVCA}.

Lastly, in \Cref{sect:OfflineVCA}, we show a polynomial-time randomized algorithm for vertex connectivity augmentation in the \emph{offline} setting, saving a factor of $\log n$ in the approximation ratio.

\subsection{Component Cuts and the Shattering Classification}
\label{ssec:component-cuts}

\paragraph{$k$-Vertex Cuts in $F$.} We use $\Ccal$ to denote the collection of $k$-vertex cuts in $F$. 
Note that the number of $k$-vertex cuts is $|\Ccal|\leq O(2^{k}\cdot n^{2})$ by \Cref{fact:num-cuts}.

\begin{definition}[Shattering and non-shattering vertex cuts]
\label{def:ShatteringVertexCuts} For each vertex cut $C\in \Ccal$, it is \emph{shattering} if $|\Gamma(F-C)|\geq k+1$. Otherwise, it is \emph{non-shattering}. \end{definition}

Let $\Ccal_{\sh}\subseteq \Ccal$ denote the collection of all shattering $k$-vertex cuts $C\in \Ccal$.
For $k\geq 2$, every shattering cut is a $k$-shredder, so $|\Ccal_{\sh}|=O(n)$ and $\Ccal_{\sh}$ can be listed in polynomial time by \Cref{fact:shredders}.
For $k=1$, every $1$-vertex cut is shattering, including the articulation points whose removal leaves exactly two components; there are at most $n$ of them.

\paragraph{Component Cuts.} For each demand pair $d=(s,t)$, a component cut $\pi$ is a tripartition $(L,C,R)$ of $V(F_{d})$ such that $|C| = r_{d}-1$, $s\in L$, $t\in R$, and there is no edge connecting $L$ and $R$ in $F_{d}$. We use $\Pi_{d}$ to denote the collection of component cuts $\pi$ of $d$. 
Note that for each non-adjacent demand pair $d$, we have $F_{d} = F$ and thus $\{C\mid \pi = (L,C,R)\in \Pi_{d}\}\subseteq \Ccal$. Similarly, we classify component cuts as either shattering or non-shattering. 
\begin{itemize}
\item For a non-adjacent demand pair $d$, a component cut $\pi = (L,C,R)\in \Pi_{d}$ is shattering if $C\in\Ccal$ is shattering, and non-shattering otherwise.
\item For an adjacent demand pair $d$, each component cut $\pi\in\Pi_{d}$ is non-shattering. Note that this is conceptually consistent with \Cref{def:ShatteringVertexCuts} by the following \Cref{ob:NonAdjacentNonShattering}.
\end{itemize}

\begin{observation}
\label{ob:NonAdjacentNonShattering}
For each component cut $\pi=(L,C,R)\in \Pi_{d}$ of an adjacent demand pair $d = (s,t)$, $F_{d}-C$ has exactly $2$ connected components.
\end{observation}
\begin{proof}
For an adjacent demand $d=(s,t)$, recall that $F_d=F-(s,t)$ and $r_d=k$.
Thus $|C|=k-1$ and $F-C$ is connected; deleting $(s,t)$ disconnects it because $(L,C,R)$ is a component cut, and therefore leaves exactly two components.
\end{proof}

\paragraph{Feasibility Criterion.} Lastly, we show that an augmentation $X\subseteq U$ is feasible if each component cut (of all demand pairs in $\Dcal$) is \emph{covered} by $X$. Here, an edge $e$ covers a component cut $\pi = (L,C,R)$ if $e$ has one endpoint in $L$ and the other in $R$. The augmentation $X$ covers $\pi$ if there exists an edge $e\in X$ covering $\pi$.

\begin{lemma}[Feasibility Criterion]
\label{lem:component-cut-exact}
Let $d=(s,t)$ be a demand pair and $X\subseteq U$ be an augmentation. Then $\kappa_{F\cup X}(s,t)\geq k+1$ if and only if every component cut $\pi = (L,C,R)\in \Pi_{d}$ is covered by $X$.
\end{lemma}
\begin{proof}
We prove both directions separately by contradiction.

\emph{($\Rightarrow$)}
Suppose $X$ fails to cover some component cut $\pi = (L,C,R)\in \Pi_{d}$.
Then no edge in $F_d\cup X$ connects $L$ and $R$, which means $C$ is a vertex cut in $F_{d}\cup X$ of size $r_{d}-1$. Therefore, $\kappa_{F_{d}\cup X}(s,t)\leq r_{d}-1$, meaning that $\kappa_{F\cup X}(s,t)\leq k$ by \Cref{ob:AdjacentReduction}, a contradiction.

\emph{($\Leftarrow$)}
Conversely, suppose $\kappa_{F\cup X}(s,t)\leq k$. Because $F$ is $k$-vertex-connected, $\kappa_{F\cup X}(s,t)\geq k$; hence the supposition gives equality. If $d$ is non-adjacent, $F_d=F$ and $r_d-1=k$. If $d$ is adjacent, deleting the edge $(s,t)$ decreases the connectivity by exactly one, so $\kappa_{F_d\cup X}(s,t)=k-1=r_d-1$. Thus in either case $\kappa_{F_{d}\cup X}(s,t)= r_{d}-1$. By definition, there is a vertex cut $C$ of size $|C| = r_{d}-1$ in $F_{d}\cup X$ such that $s$ and $t$ are disconnected in $(F_{d}\cup X)-C$. Let $L$ be the connected component of $(F_{d}\cup X) - C$ containing $s$. Then $(L,C,V(G)\setminus (L\cup C))$ is a component cut $\pi\in \Pi_{d}$ not covered by $X$, a contradiction.

\end{proof}

\subsection{Covering Non-Shattering Cuts via Hitting Set}
\label{ssec:nonshattering}

First, we show how to compute a good augmentation $X_{\HS}$ online that, at any time, covers all non-shattering component cuts of all arrived demand pairs, by solving an online hitting set instance.

The reduction itself is standard, but to ensure that the resulting hitting-set instance can be solved efficiently and yields a good approximation, we exploit the fact that the number of non-shattering component cuts is small, as shown in \Cref{lemma:NonShatteringCutsNumber}.

\begin{lemma}
\label{lemma:NonShatteringCutsNumber}
For each possible demand pair $d\in \binom{V}{2}$, the number of corresponding non-shattering component cuts is at most $O(4^{k}\cdot n^{2})$, and there is a deterministic algorithm that lists all of them in $4^{k}\cdot \poly(n)$ time.
\end{lemma}
\begin{proof}
By definition, for each component cut $\pi=(L,C,R)\in \Pi_{d}$, $C$ is a vertex cut in $F_{d}$ of size $r_{d}-1$. Since $F_{d}$ is $(r_{d}-1)$-vertex-connected (if $d$ is adjacent, this follows from \Cref{ob:MinCutSize}; if $d$ is non-adjacent, then $F_d=F$ is $k$-vertex-connected and $r_d-1=k$), the number of different $(r_{d}-1)$-vertex cuts $C$ in $F_{d}$ is bounded by $O(2^{k}\cdot n^{2})$ by \Cref{fact:num-cuts}, and there is a deterministic algorithm that lists all such vertex cuts in $2^{k}\cdot \poly(n)$ time. Therefore, it suffices to list the component cuts $(L,C,R)$ for each such vertex cut $C$.
\begin{itemize}
\item When $d$ is an adjacent demand pair, $C$ contributes at most $O(1)$ non-shattering component cuts $(L,C,R)$ by \Cref{ob:NonAdjacentNonShattering}, which can be listed in polynomial time trivially. 
\item When $d$ is a non-adjacent demand pair, $C$ contributes at most $O(2^{k})$ non-shattering component cuts $(L,C,R)$, since each of them must have a non-shattering $C$ (so $F-C$ has fewer than $k+1$ connected components) and its $L$ must be the union of a subset of connected components in $F-C$. Listing all of them can be done in $2^{k}\cdot \poly(n)$ time by computing all connected components of $F-C$ (i.e., $\Gamma(F-C)$) and then checking all bipartitions of $\Gamma(F-C)$.
\end{itemize}
In summary, the number of non-shattering component cuts of $d$ is bounded by $O(2^{k}\cdot n^{2})\cdot O(2^{k}) = O(4^{k}\cdot n^{2})$, and they can be listed in $4^{k}\cdot \poly(n)$ time.
\end{proof}

Next, we describe the online hitting set instance. The ground set is still $U$, the set of edges that can be augmented. The family of sets is $\mathcal{R}:=\{R_{\pi}\mid \text{non-shattering } \pi\in \Pi_{d},d\in\binom{V}{2}\}$, where $R_{\pi}$ is the set of $U$-edges covering $\pi$. That is, we create a set $R_{\pi}$ for every component cut of all possible demand pairs. The online sequence of sets of the hitting set instance can be naturally defined given the online sequence $\Dcal$ of demand pairs of the OVCA instance: whenever a demand pair $d\in\Dcal$ arrives, we send the sets $R_{\pi}$ of all non-shattering component cuts $\pi\in\Pi_{d}$ to the hitting set instance.

We run the deterministic online hitting set algorithm (\Cref{cor:online-hitting-set}) on the online hitting set instance, and let $X_{\HS}$ be the online solution. The following \Cref{lemma:HittingSetFeasibility} shows that $X_{\HS}$ covers all non-shattering component cuts and has low cost. It also bounds the running time. 

\begin{lemma}
\label{lemma:HittingSetFeasibility}
There is a deterministic algorithm that maintains an incremental $X_{\HS}\subseteq U$ such that at any time, $X_{\HS}$ covers every non-shattering component cut of all arrived demand pairs of the OVCA instance. Moreover, at the end, the cost of $X_{\HS}$ is at most $O((k+\log n)\cdot \log n)\cdot \OPT_{\VCA}$. The algorithm runs in $2^{O(k)}\cdot \poly(n)$ time.
\end{lemma}
\begin{proof}
Since $X_{\HS}$ is a feasible online hitting set solution and by the construction of the online hitting set instance, it is clear that $X_{\HS}$ always covers the non-shattering component cuts of the arrived demand pairs.

By \Cref{lem:component-cut-exact}, any feasible solution of the OVCA instance is also a feasible solution of the hitting set instance. Therefore, the offline optimum of the hitting set instance, denoted $\OPT_{\HS}$, is at most $\OPT_{\VCA}$. By \Cref{cor:online-hitting-set}, the competitive ratio of the online hitting set algorithm is $O(\log |U|\cdot \log|\mathcal{R}|) = O(\log n\cdot (k+\log n))$, since $|U| = O(n^{2})$ and $|\mathcal{R}| = 4^{k}\cdot \poly(n)$ by \Cref{lemma:NonShatteringCutsNumber}.
Therefore, the cost of $X_{\HS}$ is at most $O((k+\log n)\log n)\cdot\OPT_{\HS}\leq O((k+\log n)\log n)\cdot\OPT_{\VCA}$. 

Regarding the running time, constructing the hitting set instance takes $2^{O(k)}\cdot \poly(n)$ time by \Cref{lemma:NonShatteringCutsNumber}. The online hitting set algorithm runs in $\poly(|U|,|\mathcal{R}|) = 2^{O(k)}\cdot\poly(n)$ time.
\end{proof}

\subsection{Covering Shattering Cuts via Node-Weighted Steiner Forest}
\label{sect:shattering}

We cannot use the reduction to the online hitting set to cover the shattering component cuts, as their number may be as large as $2^{\Omega(n)}$. In this subsection, we show how to cover all shattering component cuts by solving a polynomial-size ONWSF instance. 

In \Cref{ssec:containment}, we introduce the Containment Lemma, a crucial structural lemma for the reduction. \Cref{ssec:shattering-reduction} and \Cref{sect:ONWSFAlgo} show the reduction and prove the following \Cref{lemma:ONWSFfeasibility}, which is one of the main technical contributions of this paper.

\begin{lemma}
\label{lemma:ONWSFfeasibility}
There is a deterministic algorithm that maintains an incremental $X_{\SF}\subseteq U$ such that at any time, $X_{\SF}$ covers all shattering component cuts of all arrived demand pairs in the OVCA instance. Moreover, at the end, the cost of $X_{\SF}$ is at most $O(\log^{2}n)\cdot \OPT_{\VCA}$ where $\OPT_{\VCA}$ is the offline optimum of the OVCA instance. The algorithm runs in $\poly(n)$ time.
\end{lemma}

\subsubsection{The Containment Lemma}
\label{ssec:containment}

The Containment Lemma below says that for a $k$-vertex cut $D$, any shattering $k$-vertex cut $C$ cannot straddle two or more connected components of $F-D$. This can be viewed as a dual form of a lemma of Jord\'an~\cite[Lemma~2.2]{Jordan95} (see \Cref{remark:DualForm}). We also show that the threshold of component number in the definition of shattering vertex cuts is essentially tight (see \Cref{remark:tight}).

\begin{lemma}[Containment Lemma]
\label{lemma:ContainmentLemma}
\label{lem:containment}
Let $F$ be a $k$-vertex-connected graph. For
\begin{itemize}
\item any shattering $k$-vertex cut $C$, and
\item any $k$-vertex cut $D$,
\end{itemize}
there is a connected component $\gamma$ of $F-D$ satisfying $C\setminus D\subseteq \gamma$, or equivalently $C\subseteq D\cup \gamma$.
\end{lemma}
\begin{proof}

We first show that, for any pair of vertices $c,c'\in C$, there exist $k+1$ internally disjoint $c$-$c'$ paths in $F$. Since $C$ is shattering, $F-C$ has at least $k+1$ connected components. For each such component $\gamma\in\Gamma(F-C)$, we construct a $c$-$c'$ path $P_{\gamma}$ as follows. \begin{itemize}
\item We take vertices $w_{\gamma},w'_{\gamma}\in \gamma$ adjacent to $c,c'\in C$, respectively. Note that $w_{\gamma},w'_{\gamma}$ must exist since every $C$-vertex is a neighbor of $\gamma$ (otherwise, the neighbor set of $\gamma$ has size $<k$, contradicting that $F$ is $k$-vertex-connected). 
\item Then $P_{\gamma}$ is the concatenation of the edge $(c,w_{\gamma})$, an arbitrary path inside $\gamma$ from $w_{\gamma}$ to $w'_{\gamma}$, and the edge $(w'_{\gamma},c')$.
\end{itemize}
Clearly, two paths $P_{\gamma}$ for two different components $\gamma$ are internally vertex disjoint.

Now, we are ready to show the original lemma. Suppose for contradiction that $C\setminus D$ intersects two different components $\gamma,\gamma'\in \Gamma(F-D)$ of $F-D$. Take arbitrary $c\in \gamma\cap C$ and $c'\in \gamma'\cap C$. We have shown that there are $k+1$ internally disjoint $c$-$c'$ paths in $F$. However, each such path must contain a vertex in $D$, contradicting that $|D| = k$.

\end{proof}

\begin{remark}
\label{remark:DualForm}
We note that Jord\'an~\cite[Lemma~2.2]{Jordan95} showed a similar lemma when $D$ is shattering instead of $C$. Therefore, our Containment Lemma can be viewed as a dual form of Jord\'an's lemma.
We point out that later we apply the Containment Lemma only when both $C$ and $D$ are shattering, so Jord\'an's lemma would also suffice for our purposes. Nevertheless, we include this dual form as it may be of independent interest.
\end{remark}

\begin{remark}
\label{remark:tight}
We also note that, for the proof of the Containment Lemma (\Cref{lemma:ContainmentLemma}) to go through, the threshold $k+1$ in \Cref{def:ShatteringVertexCuts} of shattering vertex cuts is essentially tight. In \Cref{sec:necessity}, we present a $4$-connected graph and a non-shattering $4$-component cut $C$ that straddles two sides of another minimum cut~$D$.
\end{remark}

\subsubsection{The Node-Weighted Steiner Forest Instance}
\label{ssec:shattering-reduction}

Now, we formally define the node-weighted Steiner forest instance. To avoid clutter, for any vertex cut $C\in\Ccal$ and any vertex $u\in V\setminus C$, we use $\gamma_{C}(u)$ as shorthand for $\gamma(u,F-C)$, i.e., the connected component of $F-C$ containing $u$.

\paragraph{The Auxiliary Graph $H$.} The auxiliary graph $H$ is a bipartite node-weighted graph defined as follows.

\begin{itemize}
\item (Left nodes) For each shattering vertex cut $C\in\Ccal_{\sh}$ and each component $\gamma_{C}\in \Gamma(F-C)$, create a left node with weight $0$. Without ambiguity, we still use $\gamma_{C}$ to denote this left node. 
\item (Right nodes) For each candidate edge $e\in U$, create a node with weight $c(e)$. We still use $e$ to denote this right node.
\item (Edges) For each edge $e=(u,v)\in U$ and each shattering vertex cut $C\in\Ccal_{\sh}$ such that $e$ avoids $C$ (i.e. $u,v\notin C$) and $\gamma_{C}(u)\neq \gamma_{C}(v)$ (i.e., $u$ and $v$ lie in distinct components of $F-C$), add two edges to $H$, one connecting the right node $e$ and the left node $\gamma_{C}(u)$, and the other one connecting $e$ and $\gamma_{C}(v)$.
\end{itemize}
We use $V_{L}$ to denote the set of left nodes and $V_{R} = U$ denotes the right nodes.

\paragraph{The ONWSF Demand $\mathcal{T}$.} Next, we transform the OVCA demand $\Dcal$ into the following ONWSF demand $\mathcal{T}$. For each OVCA demand pair $d=(s,t)\in \Dcal$, we define ONWSF demand pairs
\begin{equation*}
\Tcal_{d}:=\left\{
\begin{aligned}
&\{(\gamma_{C}(s),\gamma_{C}(t))\mid C\in \Ccal_{\sh}\text{  s.t. }s,t\notin C\text{ and }\gamma_{C}(s)\neq\gamma_{C}(t)\},&&\text{if $d$ is non-adjacent}\\
&\emptyset,&&\text{if $d$ is adjacent}
\end{aligned}
\right.
\end{equation*}
Then $\Tcal:=\bigcup_{d\in\Dcal}\Tcal_{d}$ simply collects all ONWSF demand pairs of all OVCA demand pairs. Naturally, whenever an OVCA demand pair $d$ arrives, we send demand pairs $\Tcal_{d}$ to the ONWSF instance.

\subsubsection{The Algorithm and Analysis: Proof of \Cref{lemma:ONWSFfeasibility}}
\label{sect:ONWSFAlgo}

We run the algorithm in \Cref{thm:onwsf} on the ONWSF instance above. Let $S\subseteq V(H)$ be the online ONWSF solution. Since every left node has zero cost, we assume that $S$ always contains $V_L$ without loss of generality. This naturally corresponds to an online augmentation $X_{\SF}\subseteq U$ by maintaining $X_{\SF}=S\setminus V_L$. The following \Cref{lemma:ONWSFfeasibility} shows that $X_{\SF}$ covers all shattering component cuts and has low cost.

\begin{lemma}
\label{lemma:FeasibleNWSFtoVCA}
Consider an OVCA demand pair $d = (s,t)\in\Dcal$. For an augmentation $X_{\SF}\subseteq U$, if all ONWSF demand pairs in $\Tcal_{d}$ are connected in $H[V_L\cup X_{\SF}]$, then $X_{\SF}$ covers all shattering component cuts of $d$.
\end{lemma}
\begin{proof}
Note that this lemma is trivial when $d = (s,t)$ is adjacent, since an adjacent $d$ has no shattering component cuts. From now on, assume $d$ is non-adjacent. First, we argue that it suffices to show that
\begin{enumerate}
\item\label{item:FeasibleNWSFtoVCA} for each shattering vertex cut $C\in\Ccal_{\sh}$ such that $s,t\notin C$ and $\gamma_{C}(s)\neq\gamma_{C}(t)$, $s$ and $t$ are connected in $(F\cup X_{\SF}) - C$.
\end{enumerate}
Once we have \Cref{item:FeasibleNWSFtoVCA}, for each shattering component cut $\pi=(L,C,R)\in\Pi_{d}$, $L$ and $R$ were previously disconnected in $F-C$ but are now connected in $(F\cup X_{\SF}) - C$, meaning that $X_{\SF}$ covers $\pi$ as desired. We devote the remainder of the proof to establishing \Cref{item:FeasibleNWSFtoVCA}.

Consider a shattering vertex cut $C\in \Ccal_{\sh}$ such that $s,t\notin C$ and $\gamma_{C}(s)\neq \gamma_{C}(t)$.
By definition, there is an ONWSF demand pair $(\gamma_{C}(s), \gamma_{C}(t))\in \Tcal_{d}$, and the lemma condition ensures that there is a path $P$ in $H[V_{L}\cup X_{\SF}]$ connecting the left nodes $\gamma_{C}(s)$ and $\gamma_{C}(t)$. Since $H$ is bipartite, the path $P$ must have the following form:
\[
(\gamma_{C}(s),~ e_{1},~ \gamma_{D_{1}},~ e_{2}, \cdots,~ e_{i},~ \gamma_{D_{i}},\cdots,~ e_{q},~ \gamma_{C}(t)),
\]
where for each $1\leq i\leq q$, $e_{i}$ is the right node corresponding to an edge $e_{i}=(u_{i},v_{i})\in X_{\SF}$, and for each $1\leq i\leq q-1$, $\gamma_{D_{i}}$ is (the left node of) some component in $\Gamma(F-D_{i})$ for some shattering vertex cut $D_{i}\in\Ccal_{\sh}$.

\begin{figure}[t]
\centering
\begin{tikzpicture}[
    scale=0.82, font=\small,
    petal/.style={draw=black!50, fill=teal!8},
    badpetal/.style={draw=black!50, fill=black!4},
    dbox/.style={draw=black!60, rounded corners=4pt, fill=white},
    hub/.style={circle, fill=teal!70!black, inner sep=0pt, minimum size=5pt},
    vtx/.style={circle, draw=black!70, fill=white, inner sep=0pt, minimum size=5pt},
    reached/.style={circle, fill=teal!70!black, draw=teal!70!black, inner sep=0pt, minimum size=5pt},
    del/.style={blue!60!black, font=\normalsize},
    purchased/.style={red!75!black, very thick},
    safe path/.style={teal!70!black, very thick},
    lab/.style={font=\scriptsize, inner sep=1.5pt}
]
\foreach \xoff in {0, 9} {
\begin{scope}[xshift=\xoff cm]
    \draw[petal] (-2.5,1.6) ellipse (1.0 and 0.55);
    \draw[petal] (-2.7,0) ellipse (1.0 and 0.55);
    \draw[petal] (-2.5,-1.6) ellipse (1.0 and 0.55);
    \draw[badpetal] (2.8,0) ellipse (1.45 and 1.9);
    \node at (3.95,1.65) {$\gamma^\star$};
    \node[del] at (2.72,0) {$\times$};
    \node[del] at (3.45,-0.5) {$\times$};
    \node[lab, blue!60!black] at (3.5,0.35) {$C\setminus D$};
    \draw[dbox] (-0.4,-1.25) rectangle (0.4,1.25);
    \foreach \p in {(-1.55,1.45),(-1.72,0),(-1.62,-1.33),(1.57,1.0)}
        \draw[black!35] (0,0.75) -- \p;
    \node at (0,1.55) {$D$};
    \node[hub, label={[lab, teal!70!black]above:$d$}] at (0,0.75) {};
    \node[hub] at (0,0.2) {};
    \node[del] at (0,-0.3) {$\times$};
    \node[del] at (0,-0.8) {$\times$};
    \node[lab, teal!70!black, anchor=west] at (0.45,0.47) {$D\setminus C$};
    \node[lab, blue!60!black, anchor=west] at (0.45,-0.55) {$C\cap D$};
\end{scope}
}

\node at (0,-3.5) {\textbf{(a)} $\gamma\neq\gamma^\star$: stay inside $\gamma$};
\node[lab, anchor=east] at (-3.55,-1.6) {$\gamma$};
\node[lab, anchor=east] at (-3.75,0) {$\gamma_i$};
\coordinate (ua) at (-3.1,-1.7);
\coordinate (ub) at (-2.0,-1.75);
\coordinate (va) at (2.2,-1.0);
\coordinate (vb) at (-2.15,0.05);
\draw[purchased] (ua) .. controls (-3.2,-3.1) and (2.2,-3.1) .. node[below, lab, red!75!black] {$e_{i-1}$} (va);
\draw[purchased] (ub) -- node[right, lab, red!75!black] {$e_i$} (vb);
\draw[safe path] (ua) to[bend left=25] (ub);
\node[reached, label={[lab]above:$u_{i-1}$}] at (ua) {};
\node[reached, label={[lab]below right:$u_i$}] at (ub) {};
\node[vtx, label={[lab]right:$v_{i-1}$}] at (va) {};
\node[vtx, label={[lab]left:$v_i$}] at (vb) {};

\begin{scope}[xshift=9cm]
\node at (0,-3.5) {\textbf{(b)} $\gamma=\gamma^\star$: detour through $W$};
\node[lab, anchor=east] at (-3.55,1.6) {$\gamma_{i-1}$};
\node[lab, anchor=east] at (-3.55,-1.6) {$\gamma_i$};
\coordinate (ua) at (2.3,1.15);
\coordinate (ub) at (2.3,-1.15);
\coordinate (va) at (-2.3,1.6);
\coordinate (vb) at (-2.3,-1.6);
\draw[purchased] (va) .. controls (-1.6,2.8) and (1.9,2.8) .. node[above, lab, red!75!black] {$e_{i-1}$} (ua);
\draw[purchased] (vb) .. controls (-1.6,-2.8) and (1.9,-2.8) .. node[below, lab, red!75!black] {$e_i$} (ub);
\draw[black!45, thick, dashed] (ua) to[bend left=40] (ub);
\draw[safe path] (va) .. controls (-2.2,1.85) and (-2.0,1.35) .. (-1.7,1.5)
    -- (0,0.75) -- (-1.7,-1.45)
    .. controls (-1.9,-1.2) and (-2.1,-1.9) .. (vb);
\node[reached] at (-1.7,1.5) {};
\node[reached] at (-1.7,-1.45) {};
\node[vtx, label={[lab]above right:$u_{i-1}$}] at (ua) {};
\node[vtx, label={[lab]below right:$u_i$}] at (ub) {};
\node[reached, label={[lab]left:$v_{i-1}$}] at (va) {};
\node[reached, label={[lab]left:$v_i$}] at (vb) {};
\end{scope}
\end{tikzpicture}
\caption{The inductive step of \Cref{claim:Induction} for an internal node $\gamma=\gamma_{D}$ of the auxiliary path, drawn in $F$ around the cut $D\neq C$.
The components of $F-D$ hang off $D$; every vertex of $D$ is adjacent to every one of them (grey lines, drawn for one vertex $d$).
By the Containment Lemma, all deleted vertices of $C$ ($\times$) lie in $D$ or in one component $\gamma^\star$.
Hence the teal region $W$ (the other components together with $D\setminus C$) survives the deletion of $C$ and is connected through $d$.
Both purchased edges $e_{i-1},e_i$ have one endpoint in $\gamma$ and the other in a different component.
The endpoint of $e_{i-1}$ lying in $W$ is reachable from $s$ (by induction, possibly using $e_{i-1}$ itself); filled dots are reachable in $(F\cup X_{\SF})-C$.
\textbf{(a)} If $\gamma\subseteq W$, walk inside $\gamma$ from $u_{i-1}$ to $u_i$.
\textbf{(b)} If $\gamma=\gamma^\star$, the route inside $\gamma^\star-C$ may be destroyed (dashed); instead go from $v_{i-1}$ to $v_i$ through $d$, using paths inside $\gamma_{i-1}$ and $\gamma_i$, never entering $\gamma^\star$.}
\label{fig:onwsf-induction-alt}
\end{figure}
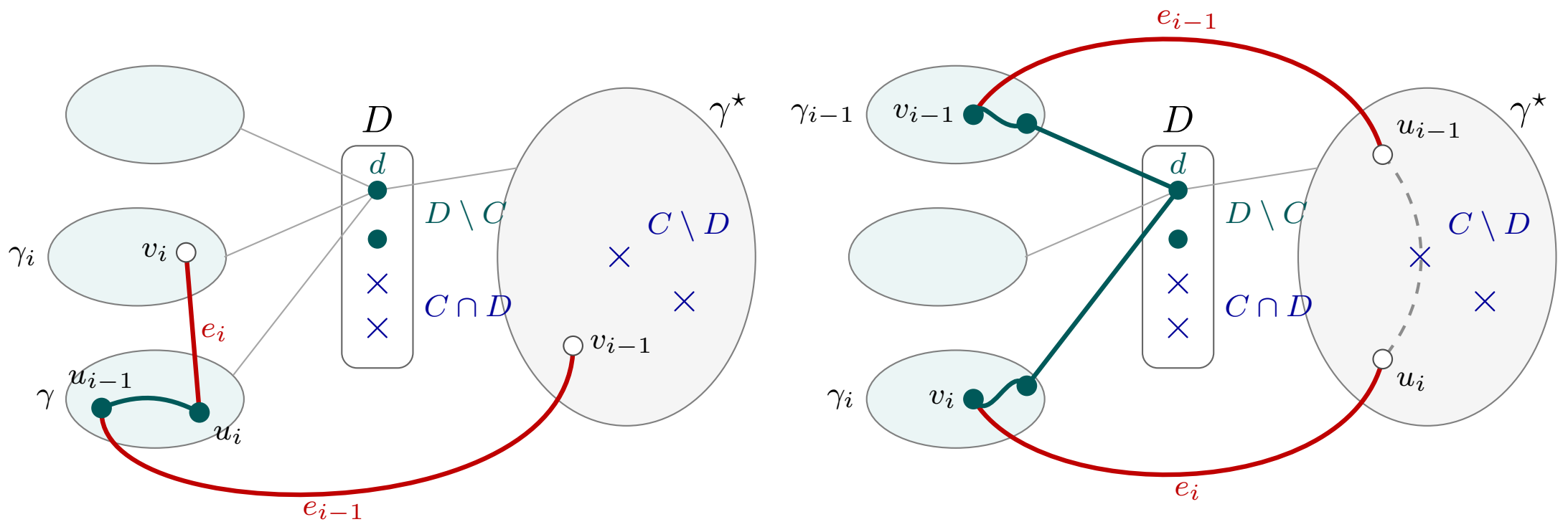

\begin{claim}
\label{claim:Induction}
For each $1\leq i\leq q$, at least one endpoint of $e_{i}$ lies outside $C$ and is connected to $s$ in $(F\cup X_{\SF})-C$.
\end{claim}
\begin{proof}
We perform induction on $i$. In the base case $i=1$, since the right node $e_{1}$ is adjacent to $\gamma_{C}(s)$ in $H$, we know that $e_{1}$ avoids $C$ and has an endpoint, say $u_{1}$, in $\gamma_{C}(s)$. Therefore, $u_{1}$ is an endpoint of $e_{1}$ outside $C$ and connected to $s$ in $(F\cup X_{\SF})-C$.

Now focus on an inductive step from $i-1$ to $i$. Consider the subpath $(e_{i-1},~\gamma_{D_{i-1}},~e_{i})$ of $P$. For brevity, we write $D$ and $\gamma$ for $D_{i-1}$ and $\gamma_{D_{i-1}}$ respectively. By the rule of adding edges in $H$, we have the following.
\begin{itemize}
\item $e_{i-1}$ avoids $D$ with one endpoint, say $u_{i-1}$, inside $\gamma$ and the other endpoint $v_{i-1}$ in some different component $\gamma_{i-1}\in \Gamma(F-D)$.
\item Similarly, $e_{i}$ avoids $D$ with one endpoint, say $u_{i}$, inside $\gamma$ and the other endpoint $v_{i}$ in some different component $\gamma_{i}\in\Gamma(F-D)$.
\end{itemize}
By the induction hypothesis, every surviving endpoint of $e_{i-1}$ is reachable from $s$: if both endpoints survive, the purchased edge $e_{i-1}$ connects them. It therefore suffices to connect a surviving endpoint of $e_i$ to a surviving endpoint of $e_{i-1}$ in $F-C$. We consider two cases; \Cref{fig:onwsf-induction-alt} illustrates the harder case $D\neq C$.

\medskip

\noindent\underline{Case $D=C$.} This case is simple. The endpoint $u_{i}$ of $e_{i}$ lies outside $C$, and it is connected to $u_{i-1}$ in $F-C$ since both $u_{i}$ and $u_{i-1}$ lie inside $\gamma\in \Gamma(F-C)$.

\medskip

\noindent\underline{Case $D\neq C$.} The argument for this case needs to exploit the containment lemma (\Cref{lemma:ContainmentLemma}). By \Cref{lemma:ContainmentLemma}, $C\setminus D$ is contained in some component $\gamma^{\star}\in \Gamma(F-D)$. Define
\[
W:= (D\setminus C) \cup \bigcup_{\gamma'\in \Gamma(F-D)\text{ s.t. }\gamma'\neq\gamma^{\star}} \gamma'.
\]
In words, $W$ is the union of $D\setminus C$ and all connected components of $F-D$ except $\gamma^{\star}$. We have:
\begin{itemize}
\item $W$ is disjoint from $C$ by definition.
\item $W$ is connected in $F-C$. This is because every vertex $v\in D\setminus C$ is a neighbor of every component $\gamma'\in \Gamma(F-D)$ (otherwise, $\gamma'$ has fewer than $k$ neighbors in $F$, contradicting that $F$ is $k$-vertex-connected). Because $C\neq D$ and $|C|=|D|=k$, the set $D\setminus C$ is nonempty, and each of its vertices joins all components included in $W$; hence $F[W]$ is connected.
\end{itemize}
Now, if $\gamma\subseteq W$, then we have $u_{i},u_{i-1}\in W$, meaning that $u_{i}$ lies outside $C$ and it is connected to $u_{i-1}$ in $F-C$. If $\gamma = \gamma^{\star}$ is not in $W$, then $\gamma_{i-1},\gamma_{i}\subseteq W$ and $v_{i},v_{i-1}\in W$, meaning that $v_{i}$ lies outside $C$ and it is connected to $v_{i-1}$ in $F-C$ (via $W$).

\end{proof}

Finally, at $i=q$, we use a similar argument. Some endpoint of $e_{q}$ lies outside $C$ and is connected to $s$ in $(F\cup X_{\SF}) - C$ by \Cref{claim:Induction}. Since the right node $e_{q}$ is adjacent to the left node $\gamma_{C}(t)$ in $H$, the edge $e_{q}$ avoids $C$, and thus both endpoints of $e_{q}$ are connected to $s$ in $(F\cup X_{\SF})-C$. Furthermore, one endpoint of $e_{q}$ falls in $\gamma_{C}(t)$, so it is connected to $t$ inside $F-C$. In summary, $s$ and $t$ are connected in $(F\cup X_{\SF})-C$ as desired.

\end{proof}

\begin{lemma}
\label{lemma:FeasibleVCAtoNWSF}
The offline optimum $\OPT_{\SF}$ for the ONWSF instance is at most $\OPT_{\VCA}$.
\end{lemma}
\begin{proof}
It suffices to show that, for any feasible offline solution $X$ to the OVCA instance, $S:=X\cup V_L$ is a feasible offline solution to the ONWSF instance. 

Consider an OVCA demand pair $d=(s,t)\in\Dcal$. Recall that, for each vertex cut $C\in\Ccal_{\sh}$ such that $s,t\notin C$ and $\gamma_{C}(s)\neq\gamma_{C}(t)$, we create an ONWSF demand pair $(\gamma_{C}(s), \gamma_{C}(t))$. Since $X$ is feasible, there is an $s$-$t$ path $P$ in $(F\cup X) - C$. Obviously\footnote{Contract every component of $F-C$ along $P$ and discard candidate edges whose endpoints lie in the same component. Every remaining $X$-edge avoids $C$ and crosses two components of $F-C$, so its right node in $H$ is adjacent to both corresponding left nodes. The resulting alternating walk contains a path from $\gamma_C(s)$ to $\gamma_C(t)$ in $H[S]$.}, this path $P$ corresponds to a path connecting $\gamma_{C}(s)$ and $\gamma_{C}(t)$ in $H[S]$.
\end{proof}

Lastly, we finish the proof of \Cref{lemma:ONWSFfeasibility}. The feasibility is guaranteed by \Cref{lemma:FeasibleNWSFtoVCA}. Combining \Cref{lemma:FeasibleVCAtoNWSF} and \Cref{thm:onwsf}, we can bound the cost of $X_{\SF}$ by $O(\log^{2}n)\cdot\OPT_{\VCA}$. 

\subsection{A Deterministic OVCA Algorithm: Proof of \Cref{thm:deterministic}}
\label{sect:DeterministicOVCA}

Given the OVCA instance, we run the algorithms in \Cref{lemma:HittingSetFeasibility} and \Cref{lemma:ONWSFfeasibility}, which maintain augmentations $X_{\HS}, X_{\SF}\subseteq U$ respectively. We maintain $X:=X_{\HS}\cup X_{\SF}$ as the OVCA solution.

To see that $X$ is feasible, \Cref{lemma:HittingSetFeasibility} and \Cref{lemma:ONWSFfeasibility} guarantee that $X_{\HS}$ (resp. $X_{\SF}$) covers all non-shattering (resp. shattering) component cuts of all arrived demand pairs. Therefore, $X$ covers all component cuts and, by \Cref{lem:component-cut-exact}, each arrived demand pair $(s,t)$ has $\kappa_{F\cup X}(s,t)\geq k+1$ as desired.

The competitive ratio is clearly $O((k + \log n)\cdot \log n)$, i.e. the sum of the competitive ratios of \Cref{lemma:HittingSetFeasibility} and \Cref{lemma:ONWSFfeasibility}. The total running time is $\poly(2^{k}, n)$ by \Cref{lemma:HittingSetFeasibility} and \Cref{lemma:ONWSFfeasibility}.

\subsection{A Polynomial-Time Randomized OVCA Algorithm}
\label{sect:RandomizedOVCA}

\begin{lemma}
\label{lemma:RandomizedOVCA}
There is a Las Vegas randomized algorithm that maintains an incremental $X\subseteq U$ such that at any time, $X$ covers every component cut of all arrived demand pairs of the OVCA instance. Moreover, at the end, the expected cost of $X$ is at most $O((k+\log n)\cdot \log n)\cdot \OPT_{\VCA}$. The algorithm runs in $\poly(n)$ time in expectation.
\end{lemma}

\subsubsection{The Online Component Cut Covering LP Instance}
\label{sect:OnlineLPInstance}

Given the OVCA instance, we define the following covering LP instance, which has a variable $x_{e}\geq 0$ for each candidate edge $e\in U$, and one covering constraint for each component cut.

\begin{equation}
\label{eq:component-cut-lp}
\begin{aligned}
\min\qquad \mathbf{c}^{\top}\mathbf{x}&\\
\text{s.t.}\qquad 
\mathbf{a}^{\top}_{\pi}\mathbf{x}&\ge1
&& \forall d\in\Dcal,\ \forall \pi=(L,C,R)\in \Pi_{d},\\
\mathbf{x} &\ge \mathbf{0}
\end{aligned}
\end{equation}
Here, $\cbf=(c(e))_{e\in U}$ denotes the cost vector of the candidate edges, and $\xbf=(x_e)_{e\in U}$ denotes the vector of LP variables. For each component cut $\pi=(L,C,R)$, $\abf_\pi=(a_{\pi,e})_{e\in U}$ denotes the coefficient vector of the corresponding constraint, where $a_{\pi,e}=1$ if $e$ covers $\pi$ (i.e., $e$ has one endpoint in $L$ and the other in $R$), and $a_{\pi,e}=0$ otherwise.

The online sequence of covering constraints is naturally defined as follows: whenever a demand pair $d\in\Dcal$ arrives, post the covering constraints of all component cuts in $\Pi_{d}$ (in an arbitrary order).

\subsubsection{A Separation Oracle}

To solve the covering LP efficiently, and round the LP solution later, we need a separation oracle described as follows. The oracle can be implemented in a standard way using a single max-flow computation. 

\begin{lemma}
\label{lemma:FractionalOracles}
Given a demand pair $d=(s,t)$ and an LP solution $\mathbf{x}=(x_e)_{e\in U}\geq \mathbf{0}$, there is a deterministic algorithm that determines if there is a component cut $\pi\in \Pi_{d}$ such that $\abf^{\top}_{\pi}\xbf<1$, i.e. the constraint corresponding to $\pi$ is violated. If such a component cut exists, the algorithm can also output one explicitly. The algorithm runs in $\poly(n)$ time.
\end{lemma}
\begin{proof}
We essentially run a single max-flow in a capacitated directed graph $G_{d,\xbf}$ via standard vertex splitting.
\begin{itemize}
\item The vertex set of $G_{d,\xbf}$ is $V(G_{d,\xbf}) = \{v^{+},v^{-}\mid v\in V\}$.
\item For each original vertex $v\in V$, add a directed edge $v^{-}\to v^{+}$ with capacity $1$ if $v\in V\setminus\{s,t\}$ and $+\infty$ otherwise.
\item For each free edge $e=(u,v)\in F_{d}$, add both directed edges $u^{+}\to v^{-}$ and $v^{+}\to u^{-}$ with capacity $+\infty$. For each candidate edge $e=(u,v)\in U$, add both directed edges with capacity $x_{e}$.
\end{itemize}
The algorithm simply computes the max-flow value from $s^{+}$ to $t^{-}$, denoted by $\lambda$. By the max-flow min-cut theorem, and the fact that each component cut $\pi=(L,C,R)\in \Pi_{d}$ has $C$ as a minimum vertex cut in $F_{d}$ (since $|C| = r_{d}-1$ and $F_{d}$ is $(r_{d}-1)$-vertex-connected), we have that
\begin{center}
there exists a component cut $\pi\in\Pi_{d}$ such that $\abf^{\top}_{\pi}\xbf<1$ if and only if $\lambda<r_{d}$.
\end{center}
Moreover, such a component cut $\pi$ can be obtained explicitly by converting a minimum $s^{+}$-$t^{-}$ cut in $G_{d,\xbf}$.
A normalized finite split-network cut corresponds to a tripartition $(L,C,R)$ of $V$ with $s\in L$, $t\in R$, and no edge of $F_d$ between $L$ and $R$ (where $C$ is the set of vertices whose split edge is cut), and has capacity $|C|+\sum_{e\in U:\,e\text{ joins }L\text{ to }R}x_e$.
Since $F_d$ is $(r_d-1)$-vertex-connected, a value below $r_d$ forces $|C|=r_d-1$ and crossing mass below $1$, giving the required component cut.
The algorithm uses only one max-flow computation, which clearly takes $\poly(n)$ time.
\end{proof}

The following is an integral version of the separation oracle, which can check the feasibility of an augmentation $X\subseteq U$.

\begin{corollary}
\label{coro:IntegralOracles}
Given a demand pair $d=(s,t)$ and an augmentation $X\subseteq U$, there is a deterministic algorithm that determines whether there is a component cut $\pi\in\Pi_{d}$ not covered by $X$. If such a component cut exists, the algorithm can also output one explicitly. The algorithm runs in $\poly(n)$ time.
\end{corollary}
\begin{proof}
Simply apply \Cref{lemma:FractionalOracles} on $d$ and $\xbf=\mathbbm{1}_X$ (namely, $x_{e}=1$ for $e\in X$ and $x_{e} = 0$ for $e\in U\setminus X$).

\end{proof}

\subsubsection{The Online Rounding: Proof of \Cref{lemma:RandomizedOVCA}}
\label{sect:OnlineRounding}

Given the OVCA instance, the algorithm first invokes two subroutines.
\begin{itemize}
\item First, it invokes \Cref{lemma:ONWSFfeasibility} to maintain an online augmentation $X_{\SF}\subseteq U$ that covers all shattering component cuts of all arrived demand pairs.
\item Also, it invokes \Cref{thm:online-covering} on the online LP instance described in \Cref{sect:OnlineLPInstance} (the required separation oracle can be implemented easily using \Cref{lemma:FractionalOracles}) to maintain a fractional LP solution $\xbf_{\LP}$. 
\end{itemize}
To obtain the desired online augmentation $X$, the algorithm performs one additional rounding step as follows. 

The algorithm maintains a growing sequence of \emph{threshold functions} $\theta_{1},\theta_{2},\ldots,\theta_{q}$ on the candidate edge set $U$. Whenever a new threshold function $\theta_i$ is created (we will see soon when this happens), its value $\theta_i(e)$ for each edge $e\in U$ is sampled independently from the uniform distribution $\Uniform[0,1]$. At any time, we maintain
\begin{equation}
\label{eq:roundingX}
X = X_{\SF}\cup\{e\in U\mid \xbf_{\LP}(e)\geq \min_{1\leq i\leq q}\{\theta_{i}(e)\}\}.
\end{equation}
(Initially $q=0$, with the minimum over an empty family interpreted as $+\infty$.)
New threshold functions are created in the following way. When a demand pair $d$ arrives, the algorithm first updates $X_{\SF}$ and $\xbf_{\LP}$ to serve it. Then, as long as the current set $X$ does not cover every component cut in $\Pi_d$ (this can be checked using \Cref{coro:IntegralOracles}), the algorithm creates a new threshold function (and updates $X$ accordingly). Note that possibly multiple threshold functions are created in one demand pair arrival.

\paragraph{Feasibility.}  Clearly, the augmentation $X$ only grows incrementally. By the algorithm, at any time, $X$ covers the component cuts of all arrived demand pairs, and thus $X$ is a feasible online augmentation by \Cref{lem:component-cut-exact}.
(The bound in \Cref{claim:NumberOfSampling} ensures almost-sure termination of the verification loop.)

\paragraph{Competitive Ratio.} Intuitively, the competitive ratio is $O((k+\log n)\cdot \log n)$ because, in expectation, each threshold function contributes $O(\log n)$, and the number of them is $O(k+\log n)$. Moreover, the contribution of $X_{\SF}$ is always bounded by $O(\log^{2}n)$. We now show a formal argument.

We first bound the expected total cost of edges in $\{e\in U\mid \xbf_{\LP}(e)\geq \min_{1\leq i\leq q}\{\theta_{i}(e)\}\}$ (i.e., the second term in \Cref{eq:roundingX}) at the end of the algorithm.
\begin{align*}
&~~~~\mathbb{E}\left[c(\{e\in U\mid \xbf_{\LP}(e)\geq \min_{1\leq i\leq q}\{\theta_{i}(e)\}\})\right]
\\
&\leq \mathbb{E}\left[\sum_{1\leq i\leq q}c(\{e\in U\mid \xbf_{\LP}(e)\geq \theta_{i}(e)\})\right]\\
&\leq \mathbb{E}\left[q\cdot (\cbf^{\top}\xbf_{\LP})\right] && \text{(by the definition of $\theta_{i}$)}\\
&\leq \mathbb{E}[q]\cdot O(\log n)\cdot \OPT_{\VCA} && \text{(by \Cref{claim:LPbound})}\\
&\leq O((k+\log n)\cdot\log n)\cdot \OPT_{\VCA} && \text{(by \Cref{claim:NumberOfSampling})}.
\end{align*}
For the second inequality, fix the oblivious demand sequence, so the final fractional vector is deterministic and independent of the threshold samples.
The event $q\geq i$ is determined before $\theta_i$ is sampled; hence the expected contribution of layer $i$ is at most $\Pr[q\geq i]\,\cbf^{\top}\xbf_{\LP}$.
Summing over $i$ gives the stated bound despite the random stopping time.
Finally, because $c(X_{\SF})\leq O(\log^{2}n)\cdot \OPT_{\VCA}$ by \Cref{lemma:ONWSFfeasibility}, we have
\[
\mathbb{E}[c(X)]\leq c(X_{\SF}) + O((k+\log n)\cdot \log n)\cdot \OPT_{\VCA}\leq O((k+\log n)\cdot \log n)\cdot \OPT_{\VCA}.
\]

\begin{claim}
\label{claim:LPbound}
$\cbf^{\top}\xbf_{\LP}\leq O(\log n)\cdot \OPT_{\VCA}$.
\end{claim}
\begin{proof}
Let $\OPT_{\LP}$ denote the offline optimum of the LP instance. By \Cref{thm:online-covering}, we have $\cbf^{\top}\xbf_{\LP}\leq O(\log |U|)\cdot \OPT_{\LP} \leq O(\log n)\cdot \OPT_{\LP}$. Also, we have $\OPT_{\LP}\leq \OPT_{\VCA}$ since a feasible offline solution to the OVCA instance is a feasible offline LP solution by \Cref{lem:component-cut-exact}.
\end{proof}
\begin{claim}
\label{claim:NumberOfSampling}
$\mathbb{E}[q]\leq O(k+\log n)$.
\end{claim}
\begin{proof}
Let $M$ be the number of distinct non-shattering component cuts appearing in the sequence, or $1$ if there are none.
For each cut $\pi$, fix the deterministic fractional vector $\xbf^\pi$ just after the LP update at the first arrival containing its constraint.
If layer $\tau$ is created, its witnessing cut is non-shattering and is missed by the first $\tau-1$ layers at the current vector, and hence also at $\xbf^\pi$ by monotonicity.
At this fixed vector, one layer misses $\pi$ with probability
\[
\prod_{e:a_{\pi,e}=1}(1-\min\{1,x_e^\pi\})\leq e^{-1},
\]
since $\abf_\pi^{\top}\xbf^\pi\geq1$.
Independence of the layers and a union bound over the fixed family of cuts give $\Pr[q\geq\tau]\leq\min\{1,Me^{-(\tau-1)}\}$, even though the witnessing cut is chosen adaptively.
Therefore
\[
\mathbb{E}[q] = \sum_{\tau\geq 1}\Pr[q\geq \tau]\leq \ln M + O(1) \leq O(k + \log n).
\]
The last inequality is because the number of non-shattering component cuts over all possible demand pairs is at most $\poly(2^{k},n)$ by \Cref{lemma:NonShatteringCutsNumber}.
\end{proof}

\paragraph{Running Time.} Maintaining $X_{\SF}$ online takes $\poly(n)$ time by \Cref{lemma:ONWSFfeasibility}. Maintaining $\xbf_{\LP}$ takes $\poly(n)$ time by \Cref{thm:online-covering} and because the separation oracle in \Cref{lemma:FractionalOracles} takes $\poly(n)$ time for each query. Finally, the online rounding step takes $\poly(n)$ time in expectation, since \Cref{coro:IntegralOracles} takes $\poly(n)$ time to verify the feasibility.

\subsection{A Polynomial-Time Offline VCA Algorithm}
\label{sect:OfflineVCA}

\begin{theorem}
\label{thm:OfflineVCA}
There is a Las Vegas randomized algorithm that,
given an offline VCA instance,
computes an augmentation $X\subseteq U$ such that $\kappa_{F\cup X}(s,t)\geq k+1$ for every demand pair $(s,t)\in\Dcal$.
The expected cost of $X$ is at most $O(k + \log n)\cdot \OPT_{\VCA}$. The algorithm runs in $\poly(n)$ time in expectation.
\end{theorem}
\begin{proof}
The algorithm is essentially the same as the polynomial-time online algorithm, except that the online subroutines are replaced by their offline counterparts. The algorithm is as follows.
\begin{itemize}
\item First, we compute an (offline) augmentation $X_{\SF}\subseteq U$ that covers all shattering component cuts, by applying the offline node-weighted Steiner forest algorithm (\Cref{thm:klein-ravi}) on the ONWSF instance in \Cref{ssec:shattering-reduction}. 
\item Second, solve the covering LP in \Cref{sect:OnlineLPInstance} offline using the ellipsoid method with the separation oracle in \Cref{lemma:FractionalOracles}, which gives a fractional augmentation $\xbf_{\LP}$ that covers all component cuts.
\item Lastly, apply an offline version of the online rounding in \Cref{sect:OnlineRounding} to $X_{\SF}$ and $\xbf_{\LP}$, which gives an augmentation $X\subseteq U$ that covers all component cuts.
(Sample independent threshold layers one at a time, checking all demands with \Cref{coro:IntegralOracles} after each layer, and stop only when all pass.
The same tail bound gives $\mathbb E[q]=O(k+\log n)$, so this is Las Vegas with expected polynomial running time.)
\end{itemize}

The arguments for feasibility, approximation guarantee, and running time are essentially the same as those for the online algorithm. For clarity, we sketch the approximation analysis below. 

Since \Cref{thm:klein-ravi} is $O(\log n)$-approximate, we can show that $c(X_{\SF})\leq O(\log n)\cdot \OPT_{\VCA}$ following the proof of \Cref{lemma:ONWSFfeasibility}. The second step solves the LP exactly, so $\cbf^{\top}\xbf_{\LP} = \OPT_{\LP}$. Lastly, following the analysis in \Cref{sect:OnlineRounding}, we have $\mathbb{E}[c(X)]\leq c(X_{\SF}) + \mathbb{E}[q\cdot (\cbf^{\top}\xbf_{\LP})]\leq O(k+\log n)\cdot \OPT_{\VCA}$.

\end{proof}

\section{Online $\{1,2\}$-VC-SNDP}\label{sec:1-2-online}

In this section, we will present an online algorithm for general vertex-connectivity SNDP with connectivity requirements at most $2$.

Observe that, when the input graph $G$ contains a free spanning tree $F$, this is exactly the OVCA problem when $k=1$. To remove the free-tree assumption, we borrow ideas from the online EC-SNDP algorithm by Gupta, Krishnaswamy, and Ravi~\cite{GuptaKR09}, which exploits a probabilistic low-stretch spanning tree sampler (\Cref{thm:LowStretchTree}).

\begin{theorem}[Probabilistic low-stretch spanning tree {\cite{AbrahamBN08}}]
\label{thm:LowStretchTree}
Let \(G=(V,E)\) be an \(n\)-vertex connected graph with positive edge
lengths/costs. There is a randomized polynomial-time algorithm that outputs a
spanning tree \(T\subseteq G\), with each tree edge retaining its original
length, such that, for every \(u,v\in V\),
\[
    \dist_G(u,v)\le \dist_T(u,v)
    \quad\text{and}\quad
    \mathbb{E}[\dist_T(u,v)]\le \widetilde O(\log n)\,\dist_G(u,v),
\]
where $\dist_{G}(u,v)$ and $\dist_{T}(u,v)$ denote the $u$-$v$ distance in $G$ and $T$ respectively. 
\end{theorem}

The benefit of considering such a low-stretch spanning tree $T$ is that, whenever we plan to purchase an edge $e$, it is affordable to simultaneously purchase the whole fundamental cycle of $e$. This makes the tree $T$ effectively free without buying the whole tree in advance.

This is illustrated by the following \Cref{lemma:EffectivelyFree}: if an augmentation $X\subseteq E(G)\setminus E(T)$ makes a demand pair $(s,t)$ $2$-vertex-connected in $T\cup X$, then $(s,t)$ is still $2$-vertex-connected in the union of the $s$-$t$ tree path and the fundamental cycles of all $X$-edges.

\begin{lemma}
\label{lemma:EffectivelyFree}
Let $T$ be any spanning tree of $G$, and let $T[u,v]$ denote the $u$-$v$ tree path for each vertex pair $(u,v)$. Consider an augmentation $X\subseteq E(G)\setminus E(T)$. Then, for each pair of distinct vertices $s,t\in V(G)$, $s$ and $t$ are $2$-vertex-connected in $T\cup X$ if and only if they are $2$-vertex-connected in the subgraph
\[
Q_{s,t}:= T[s,t]\cup \bigcup_{e=(u,v)\in X}(\{e\}\cup T[u,v]).
\]
\end{lemma}
\begin{proof}
We only show the direction $(\Rightarrow)$, since the other direction is trivial.
Every simple $s$-$t$ path $P$ in $T\cup X$ lies in $Q_{s,t}$: if $f\in P\cap E(T)$ is not in $T[s,t]$, then $s,t$ lie on the same side of the fundamental cut of $f$, so $P$ must cross it again using an $X$-edge $e=(u,v)$; hence $f\in T[u,v]\subseteq Q_{s,t}$.
Thus two internally vertex-disjoint paths in $T\cup X$ remain present in $Q_{s,t}$, including when $s,t$ are adjacent.
\end{proof}

Now we are ready to show the online $\{1,2\}$-VC-SNDP algorithm.

\begin{theorem}
\label{thm:VCSNDPfull}
Given an undirected graph $G = (V,E)$ with edge costs, and an online sequence of demand pairs $(s,t)\in \binom{V}{2}$, each with a connectivity requirement $r_{s,t}\in\{1,2\}$, there is a randomized algorithm that maintains an incremental edge set $S\subseteq E$ such that at any time, $\kappa_{(V,S)}(s,t)\geq r_{s,t}$ for each arrived $(s,t)$. Moreover, at the end, the expected cost of $S$ is at most $\widetilde{O}(\log^{3}n)\cdot\OPT$. The algorithm runs in polynomial time.
\end{theorem}
\begin{proof}
The algorithm is as follows.
\begin{itemize}
\item Initially, compute a low-stretch spanning tree $T$ using \Cref{thm:LowStretchTree}.\footnote{\Cref{thm:LowStretchTree} assumes that $G$ is connected and has positive costs. We may assume that $G$ is connected, as otherwise we can run the algorithm independently in each connected component (a demand pair crossing components is infeasible). To handle zero costs, we use the following standard preprocessing: choose a zero-cost spanning forest inside every connected component of the zero-cost subgraph, contract those components, sample the low-stretch tree on the positive-cost quotient, and expand the zero-cost forests. This preserves the stated stretch inequality for the original nonnegative costs.}
\item When processing the online sequence of demand pairs, run the deterministic OVCA algorithm in \Cref{thm:deterministic} with $k=1$, $T$ as the free graph and those demand pairs with connectivity requirement $2$, where each non-tree edge $e=(u,v)$ is assigned the inflated cost $c_T(e):=c(e)+c(T[u,v])$. Let $X\subseteq E\setminus E(T)$ be the online augmentation. This way, the cost of buying the fundamental cycles of all edges in $X$ is at most the cost $c_T(X)$ charged by the OVCA algorithm.
\item Always maintain
\[
S:=\bigcup_{\substack{\text{arrived demand}\\\text{ pairs $(s,t)$}}} T[s,t]\cup \bigcup_{e=(u,v)\in X}(\{e\}\cup T[u,v])
\]
as the online VC-SNDP solution.
\end{itemize}

The feasibility of $S$ follows directly: requirement-$1$ pairs are connected by their tree paths, while \Cref{lemma:EffectivelyFree} transfers the OVCA guarantee for requirement-$2$ pairs to the maintained subgraph $S$. The running time is clearly $\poly(n)$ by \Cref{thm:deterministic}. We focus on analysing the competitive ratio. We bound the cost of two terms of $S$ separately.

Consider the first term, i.e., the union of $T[s,t]$ over all arrived demand pairs $(s,t)$, denoted by $S_{1}$. Let $S^{\star}$ be an optimal offline solution. Since $T$ is a low-stretch spanning tree, the union of fundamental cycles of $S^{\star}$-edges, i.e.,
\[
\hat{S}^{\star}:=\bigcup_{e=(u,v)\in S^{\star}}(\{e\}\cup T[u,v])
\]
has expected cost at most $\widetilde{O}(\log n)\cdot \OPT$. Observe that $S_{1}\subseteq \hat{S}^{\star}$ since every demand pair $(s,t)$ is connected in $(V,S^{\star})$. Indeed, removing a tree edge $f\in T[s,t]$ separates $s$ from $t$ in $T$, so any $s$-$t$ path in $S^{\star}$ contains an edge $e=(u,v)$ crossing this fundamental cut, and then $f\in T[u,v]$. Thus, the expected cost of $S_{1}$ is at most $\widetilde{O}(\log n)\cdot \OPT$.

Consider the second term, i.e. the union of $\{e\}\cup T[u,v]$ over all $X$-edges $e$, denoted by $S_{2}$. Note that $X$ depends on the sampled tree $T$, so we cannot directly apply the low-stretch guarantee to the edges of $X$. Instead, we compare against $S^{\star}$, which is fixed independently of $T$ (as the demand sequence is fixed in advance). Let $\beta=O(\log^{2}n)$ be the competitive ratio of \Cref{thm:deterministic} for $k=1$. For every sampled $T$, the set $S^{\star}\setminus E(T)$ is a feasible solution of the OVCA instance solved in the second step, since $T\cup(S^{\star}\setminus E(T))\supseteq S^{\star}$. Hence,
\[
c(S_2)\leq c_T(X)\leq\beta\cdot c_T(S^{\star}\setminus E(T)).
\]
Only now do we take the expectation over $T$. The low-stretch guarantee and $\dist_G(u,v)\leq c(e)$ for each $e=(u,v)\in S^{\star}$ imply
\[
\mathbb{E}_T[c_T(S^{\star}\setminus E(T))]\leq\sum_{e=(u,v)\in S^{\star}}\bigl(c(e)+\mathbb{E}_T[c(T[u,v])]\bigr)\leq\widetilde O(\log n)\cdot\OPT.
\]
Therefore, $\mathbb{E}[c(S_2)]\leq\widetilde O(\log^{3}n)\cdot\OPT$.

In summary, at the end the expected cost of $S$ is at most $\widetilde{O}(\log^{3}n)\cdot\OPT$.
\end{proof}

\begin{remark}
\label{rem:offline-012}
The offline version of the $\{1,2\}$-VC-SNDP problem has a deterministic $2$-approximation~\cite{Fleischer01VC012,FleischerJW06}.
(Online, even the $r=1$ subcase has the classical $\Omega(\log n)$ lower bound from online Steiner tree~\cite{ImaseW91}.)
\end{remark}

\section{Lower Bounds for Vertex-Connectivity Augmentation}
\label{sec:hardness}

In this section
we record the consequences of applying the framework of Lando and Nutov~\cite{LandoN09}
to obtain a reduction
from
\emph{Directed Steiner Network} (DSN), also known as \emph{Directed Steiner Forest},
to
vertex-connectivity augmentation.

In the DSN problem, the input is a directed graph $J = (V(J), E(J))$ with \emph{edge} costs, and a collection of directed demand pairs $\mathcal{R} = \{(p_{1},q_{1}),(p_{2},q_{2}),...\}$. The goal is to find a minimum-cost edge set $S\subseteq E(J)$ such that in the subgraph $(V(J),S)$, $p$ can reach $q$ for each $(p,q)\in\mathcal{R}$.
In the online version of DSN, the demand pairs arrive online.

\begin{theorem}[Reduction from DSN to VCA~\cite{LandoN09}]
\label{thm:ReductionDSNtoVCA}
Given a \DSN instance $(J,\mathcal{R})$ on $\hat{n}$ vertices and $D$ demand pairs, there exists a VCA instance $(G,F,k,\mathcal{D})$ on $n=2\hat{n}$ vertices with connectivity parameter $k=\hat{n}$ and $D$ demand pairs, such that the two instances have the same optimum value. Moreover, the VCA instance can be constructed in $\poly(n)$ time. The reduction also applies online.
\end{theorem}
We give a proof for completeness.
\begin{proof}
We first construct the VCA instance $(G,F,k,\mathcal{D})$. The undirected graph $G$ is as follows.\footnote{We may assume without loss of generality that every demand pair $(p,q)$ has $p\neq q$, since self-demands are already feasible and can be discarded. We also remove self-loops from $J$ and, among parallel directed edges with the same ordered endpoints, retain only the cheapest one. These operations preserve the optimum for every collection of demands and ensure that the candidate graph constructed below is simple.}
\begin{itemize}
\item The vertices are $V(G):=\{v_{l},v_{r}\mid v\in V(J)\}$, i.e., for each vertex $v\in V(J)$, create two copies $v_{l},v_{r}$ in $G$. Let $V_{L} = \{v_{l}\mid v\in V(J)\}$ and $V_{R} = \{v_{r}\mid v\in V(J)\}$.
\item The free graph $F$ is made up of a clique on $V_{L}$, a clique on $V_{R}$, and a matching that connects $(v_{l},v_{r})$ for each $v\in V(J)$.
\item For each directed edge $(u,v)\in E(J)$, add an undirected edge $(u_{l},v_{r})$ to $G$ with the same cost. The candidate edge set $U$ consists of all these $(u_{l},v_{r})$.
\end{itemize}
The connectivity parameter is $k=\hat{n}$, and the demand set $\Dcal$ of the VCA instance is
\[
\Dcal = \{(p_{l},q_{r})\mid (p,q)\in \mathcal{R}\}.
\]
Clearly, this mapping can also be performed online.

First, this is a valid VCA instance because the free graph $F$ is $k$-vertex-connected. Indeed, after removing any set of fewer than $k=\hat n$ vertices, the remaining graph consists of two non-empty cliques joined by at least one matching edge, and is therefore connected.

It remains to show that both instances have the same optimum value.
Since edges $E(J)$ are in one-to-one correspondence with candidate edges $U$, a DSN solution $S\subseteq E(J)$ naturally corresponds to a VCA solution $X\subseteq U$. Next, it suffices to show that $S$ is feasible if and only if $X$ is feasible. That is, for each demand pair $(p,q)\in \mathcal{R}$,
\begin{center}
$p$ can reach $q$ in $(V(J),S)$ ~$\Longleftrightarrow$~ $\kappa_{F\cup X}(p_{l},q_{r})\geq k+1$.
\end{center}
Without loss of generality, assume $p$ and $q$ are not the same vertex in $V(J)$, and thus $(p_{l},q_{r})\in \mathcal{D}$ is non-adjacent in $F$.

\medskip

\noindent{\textit{Direction} $(\Rightarrow)$.} Suppose for contradiction that $\kappa_{F\cup X}(p_{l},q_{r})\leq k = \hat{n}$. Then observe that, in graph $F$, every component cut $\pi=(L,C,R)$ of $d=(p_{l},q_{r})$ is of the form, for some $A\subseteq V(J)$ such that $p\in A$ and $q\notin A$,
\begin{equation}
\label{eq:ComponentCutofA}
\begin{aligned}
L&= \{v_{l}\mid v\in A\},\\
C&= \{v_{l}\mid v\in V(J)\setminus A\}\cup \{v_{r}\mid v\in A\},\\
R&= \{v_{r}\mid v\in V(J)\setminus A\}.
\end{aligned}
\end{equation}
Indeed, since $V_L$ and $V_R$ are cliques with $p_l\in L$ and $q_r\in R$, every left copy outside $C$ lies in $L$ and every right copy outside $C$ lies in $R$. Moreover, $C$ contains at least one endpoint of each matching edge $v_lv_r$, since otherwise that edge would connect $L$ and $R$. As $|C|=\hat n$, it contains exactly one endpoint of every matching edge, which gives the shown form for $A=\{v\in V(J)\mid v_l\notin C\}$.

By \Cref{lem:component-cut-exact}, $\kappa_{F\cup X}(p_{l},q_{r})\leq k$ means there exists such a component cut $(L,C,R)$ such that no $X$-edge connects $L$ and $R$. Then by the construction of the VCA instance, there is no directed edge in $S$ from $A$ to $V(J)\setminus A$, contradicting that $p$ can reach $q$ in $(V(J),S)$.

\medskip

\noindent{\textit{Direction} $(\Leftarrow)$.} Suppose $\kappa_{F\cup X}(p_{l},q_{r})\geq k+1$ and, for contradiction, that $p$ cannot reach $q$ in $(V(J),S)$. Then there exists a set $A\subseteq V(J)$ such that $p\in A$, $q\in V(J)\setminus A$, and there is no $S$-edge from $A$ to $V(J)\setminus A$. Observe that $(L,C,R)$ defined in \Cref{eq:ComponentCutofA} forms a component cut of $(p_{l},q_{r})$. Moreover, there is no $X$-edge connecting $L$ and $R$ by the construction of the VCA instance. By \Cref{lem:component-cut-exact}, this gives $\kappa_{F\cup X}(p_{l},q_{r})\leq k$, a contradiction.
\end{proof}

The following hardness results are known for DSN.

\begin{theorem}[\cite{DodisK99}, Theorem 4.2]
\label{thm:DK99}
For any constant $\epsilon>0$, 
\begin{itemize}
\item there is no polynomial-time deterministic algorithm for DSN with approximation $O(2^{\log^{1-\epsilon} n})$ unless $\mathrm{NP}\subseteq\mathrm{DTIME}(n^{\polylog n})$,
\item there is no polynomial-time randomized algorithm for DSN with constant success probability and approximation $O(2^{\log^{1-\epsilon} n})$ unless $\mathrm{NP}\subseteq\mathrm{BPTIME}(n^{\polylog n})$
\end{itemize}
\end{theorem}

\begin{theorem}[\cite{HLMT27}] \label{thm:HLMT}
For some constant $c > 0$,
the competitive ratio of any online algorithm for DSN
is at least $\exp(c \sqrt{\log n})$.
This holds even for randomized algorithms against an oblivious adversary.
\end{theorem}

Combining the reduction (\Cref{thm:ReductionDSNtoVCA}) with \Cref{thm:DK99} and \Cref{thm:HLMT} immediately gives the hardness results in \Cref{thm:hardness-intro}, which we restate for convenience.

\hardnessconsequencesthm*

Moreover, we emphasize that obtaining a polynomial-time algorithm for VCA with approximation $k^{\alpha}\polylog(n)$ for appropriate $\alpha<1$ would advance the approximation frontier for Directed Steiner Network, as shown in the following \Cref{coro:VCAtoDSN}. In the offline setting, the state-of-the-art polynomial-time approximation ratio for DSN is $O(n^{2/3+\epsilon})$~\cite{BermanBMRY13} (and $\widetilde O(n^{4/7+\varepsilon})$ for uniform costs~\cite{AbboudB18}), and $O(D^{1/2+\epsilon})$~\cite{ChekuriEGS11,FeldmanKN12}. In the online setting, the state-of-the-art polynomial-time competitive ratio for online DSN is $O(D^{1/2+\epsilon})$~\cite{ChakrabartyEKP18} for general costs and $\widetilde{O}(n^{3/5})$ for uniform costs~\cite{GrigorescuLQ21,DBLP:conf/soda/BodwinL25}.
The following is a corollary of \Cref{thm:ReductionDSNtoVCA}.

\begin{corollary} 
\label{coro:VCAtoDSN}
For an arbitrary constant $\alpha>0$, we have the following.
\begin{itemize}
\item
In the offline setting, any polynomial-time algorithm for VCA with approximation $k^{\alpha}\cdot \polylog n$ implies a polynomial-time $\widetilde{O}(n^{\alpha})$-approximation algorithm for DSN. 
\item Similarly, in the online setting, any polynomial-time online algorithm for OVCA with competitive ratio $k^{\alpha}\cdot\polylog n$ implies a polynomial-time online algorithm for online DSN with competitive ratio $\widetilde{O}(n^{\alpha})$.
\end{itemize}
\end{corollary}

Lastly, we note that, in the online setting, we cannot hope for an OVCA algorithm with a competitive ratio depending only on $k$ (e.g., $O(k)$-competitive), by the following \Cref{prop:log-lb}.

\begin{proposition}
\label{prop:log-lb}
Every online OVCA algorithm has a competitive ratio $\Omega(\log n)$.
\end{proposition}
\begin{proof}
We can reduce an online Steiner forest instance on $(V,E)$ to an OVCA instance with $k=1$ as follows. Add a virtual vertex $h$ and star edges $\{(h,v)\mid v\in V\}$. Let the free graph $F$ be the star, and make the candidate edges $U$ exactly $E$ with the same costs. Map each Steiner demand pair $(s,t)$ to the OVCA demand pair $(s,t)$.

Then for any augmentation $X\subseteq U$, we have $\kappa_{F\cup X}(s,t)\ge2$ if and only if $s,t$ are connected in $(V,X)$:
a connecting path in $(V,X)$ plus the path $(s,h,t)$ gives two internally disjoint paths;
conversely two disjoint paths use $h$ at most once, so one avoids $h$ and lies in $X$.

The statement now follows from the
lower bound $\Omega(\log n)$ on the competitive ratio
of
online Steiner forest (even online Steiner tree)~\cite{ImaseW91}.
\end{proof}

\paperopenquestions
\paperacknowledgements
\appendix

\section{Necessity of Shattering}

\label{sec:necessity}

In this section, we show that, to obtain the Containment Lemma, the component-number threshold $k+1$ in \Cref{def:ShatteringVertexCuts} of shattering vertex cuts cannot be weakened. For example, we cannot only work with shredders (vertex cuts with at least 3 components).
We exhibit a $4$-connected graph with a non-shattering shredder~$C$ (four components, but $4<k+1=5$) and a second minimum cut~$D$ that $C$ straddles, so that the Containment Lemma fails. Moreover, we show that on this instance, the ONWSF reduction, applied indiscriminately to all minimum cuts, declares a non-solution feasible.

\begin{figure}[t]
\centering
\begin{tikzpicture}[
    vertex/.style={circle, draw, fill=white, minimum size=6pt, inner sep=1pt, font=\scriptsize},
    wvertex/.style={circle, draw, fill=gray!30, minimum size=8pt, inner sep=1pt, font=\scriptsize},
    purchased/.style={thick, red, dashed},
    scale=0.9
]
\foreach \i/\xoff in {1/0, 2/3.5, 3/7, 4/10.5} {
    \node[vertex] (v\i1) at (\xoff, 1.2) {$v_{\i 1}$};
    \node[vertex] (v\i2) at (\xoff+1.5, 1.2) {$v_{\i 2}$};
    \node[vertex] (v\i3) at (\xoff, -0.3) {$v_{\i 3}$};
    \node[vertex] (v\i4) at (\xoff+1.5, -0.3) {$v_{\i 4}$};
    \draw (v\i1) -- (v\i2);
    \draw (v\i1) -- (v\i3);
    \draw (v\i1) -- (v\i4);
    \draw (v\i2) -- (v\i3);
    \draw (v\i2) -- (v\i4);
    \draw (v\i3) -- (v\i4);
    \node[above=2pt] at (\xoff+0.75, 1.4) {$K_4^{(\i)}$};
}
\foreach \j/\yoff in {1/3, 2/3.8, 3/4.6, 4/5.4} {
    \node[wvertex] (w\j) at (6, \yoff) {$w_{\j}$};
}
\foreach \j in {1,2,3,4} {
    \foreach \i/\xoff in {1/0, 2/3.5, 3/7, 4/10.5} {
        \ifnum\i\j=44\else\ifnum\i\j=13\else\draw[thin, gray] (w\j) -- (v\i\j);\fi\fi
    }
}
\draw[thin, gray] (w4) to[bend left=16] (v44);
\draw[thin, gray] (w3) to[bend left=10] (v13);
\draw[purchased] (v11) to[bend right=30] (v21);
\draw[purchased] (v21) to[out=105, in=200] (w3);
\draw[purchased] (w3) to[out=-20, in=90] (v31);
\draw[purchased] (v31) to[bend right=30] (v41);
\foreach \i in {1,2,3,4} {
    \draw[thick, green!60!black, fill=green!20, opacity=0.35] (v\i3) circle (0.33);
}
\node[left=1pt, font=\small] at (v11.west) {$s$};
\node[left=1pt, yshift=-1pt, font=\small] at (v41.west) {$t$};
\node[green!50!black, font=\small] at (5.7, -1.0) {$D = N_F(w_3) = \{v_{13}, v_{23}, v_{33}, v_{43}\}$};
\draw[thick, blue, dashed, rounded corners=5pt] (4.8, 2.5) rectangle (7.2, 6);
\node[blue, font=\small] at (6, 6.3) {$C = \{w_1, w_2, w_3, w_4\}$};
\end{tikzpicture}
\caption{The $k=4$ counterexample.
The free graph~$F$ has four disjoint $K_4$ copies and four bridge vertices $w_j$, with $w_j$ adjacent to $v_{ij}$ for all~$i$.
The blue cut $C=\{w_1,w_2,w_3,w_4\}$ is a non-shattering shredder ($|\Gamma(F-C)|=4<k{+}1=5$).
The green cut $D=N_F(w_3)$ isolates $w_3$, so $F-D$ has the two components $\{w_3\}$ and the rest; thus $C$ has $w_3$ on one side and $w_1,w_2,w_4$ on the other, and the Containment Lemma fails.
The red augmentation edges form a legal path in the auxiliary graph for the demand $(s,t)=(v_{11},v_{41})$, yet the corresponding $s$--$t$ walk passes through $w_3\in C$ and is destroyed by deleting~$C$.}
\label{fig:counterexample}
\end{figure}
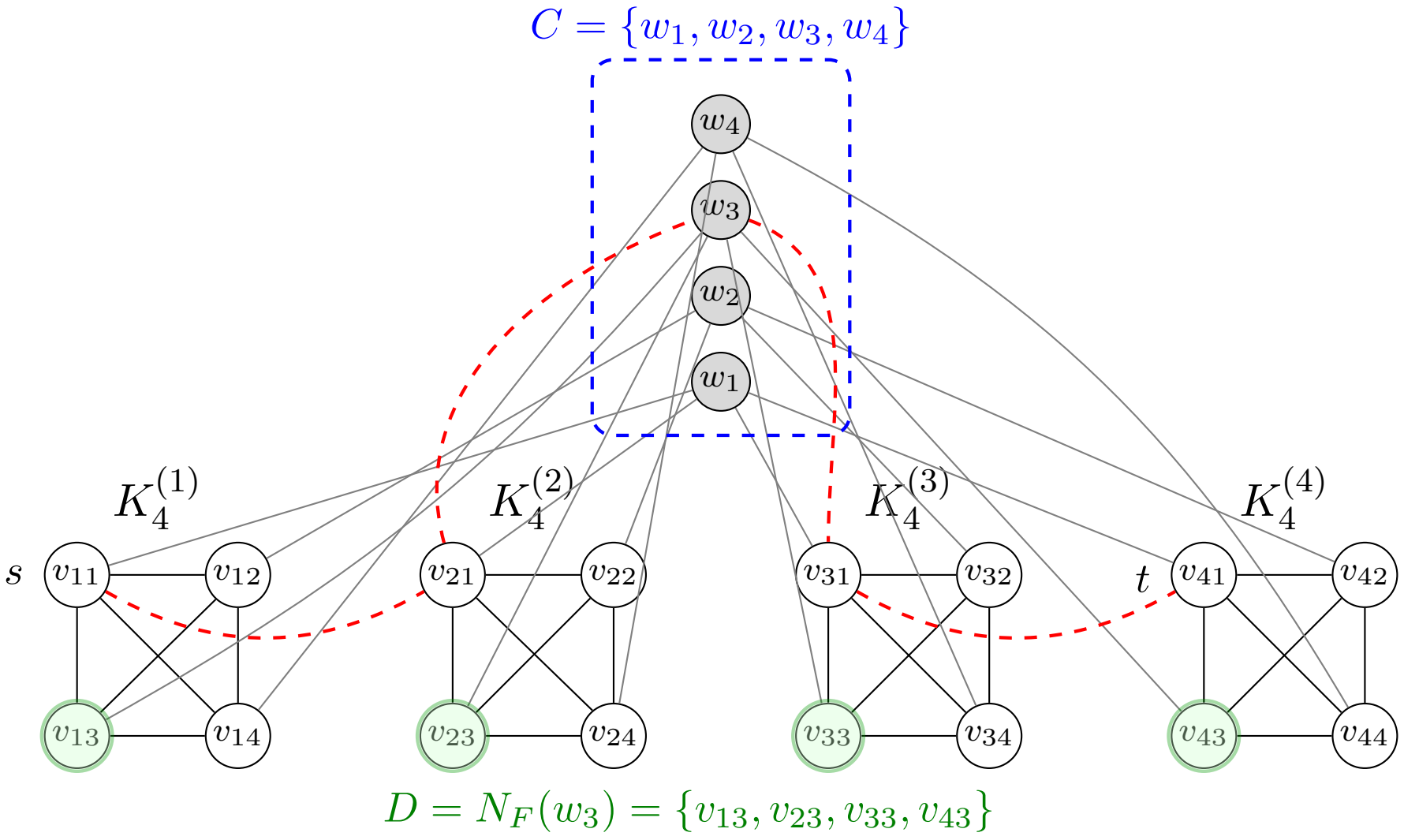

\paragraph{Construction.}
Let $k=4$.
The free graph~$F$ has four disjoint copies of $K_4$, with vertex set $\{v_{ij}:i,j\in\{1,2,3,4\}\}$ where $K_4^{(i)}=\{v_{i1},v_{i2},v_{i3},v_{i4}\}$, together with four bridge vertices $w_1,\dots,w_4$, where $w_j$ is adjacent to $v_{ij}$ for every~$i$.
The candidate edge set $U$ contains every non-free edge, with positive cost.

\begin{claim}
$F$ is $4$-connected.
\end{claim}
\begin{proof}
When deleting any set $Z$ with $|Z|\le3$, at least one bridge $w_j$ survives.
First, every copy $K_4^{(i)}-Z$ still has a vertex adjacent to a surviving bridge: otherwise, for every surviving bridge~$w_j$ the corresponding neighbor $v_{ij}$ would be deleted, and accounting for the deleted bridges this already forces $\ge4$ deletions.
Second, any two surviving bridges $w_p,w_q$ lie in one component: if no copy $K_4^{(i)}-Z$ contained both $v_{ip}$ and $v_{iq}$, then each of the four copies would lose one of them, again $\ge4$ deletions.
Hence all surviving bridges, and through them all surviving copies, lie in one component, so $F-Z$ is connected.
\end{proof}

\paragraph{Failure of containment.}
Let $C=\{w_1,w_2,w_3,w_4\}$.
Then $F-C$ is exactly the four $K_4$ copies, so $C$ is a minimum cut with four components; since $4<k+1=5$, it is non-shattering (though a shredder).
Let $D=N_F(w_3)=\{v_{13},v_{23},v_{33},v_{43}\}$.
Deleting~$D$ isolates~$w_3$ (all its neighbors are gone) while the other $15$ vertices stay connected (each $K_4^{(i)}-D$ is nonempty and connected, and $w_1,w_2,w_4$ tie the copies together).
So $F-D$ has the two components $\gamma_{\mathrm{small}}=\{w_3\}$ and $\gamma_{\mathrm{big}}=V(F)\setminus(D\cup\{w_3\})$.
Now $D\cup \gamma_{\mathrm{small}}$ contains only $w_3$ from~$C$, while $D\cup \gamma_{\mathrm{big}}$ contains $w_1,w_2,w_4$ but not~$w_3$; hence $C\not\subseteq D\cup \gamma$ for either component~$\gamma$.
The Containment Lemma fails, exactly because $C$ is not shattering.

\paragraph{Failure of the reduction.}
Take the demand $(s,t)=(v_{11},v_{41})$ and consider applying the Steiner-forest construction of \Cref{ssec:shattering-reduction} to \emph{every} minimum cut, ignoring the shattering classification.
The cut $C$ contributes the component-nodes of its four $K_4$ components, while each of $N_F(v_{21})$, $D=N_F(w_3)$, and $N_F(v_{31})$ isolates one vertex and contributes the corresponding component-node. The four candidate edges
\[
e_1=(v_{11},v_{21}),\quad e_2=(v_{21},w_3),\quad e_3=(w_3,v_{31}),\quad e_4=(v_{31},v_{41})
\]
yield the legal auxiliary path
\[
\gamma_{C}(s)-e_1-\gamma_{N_F(v_{21})}(v_{21})-e_2-\gamma_{D}(w_3)-e_3-\gamma_{N_F(v_{31})}(v_{31})-e_4-\gamma_{C}(t).
\]
If these four edges are cheap and all others expensive, the Steiner-forest algorithm may purchase exactly $X=\{e_1,e_2,e_3,e_4\}$.
But the only $s$--$t$ walk they provide is $v_{11}-v_{21}-w_3-v_{31}-v_{41}$, which runs through $w_3\in C$; after deleting~$C$ it is gone, so $(F\cup X)-C$ does not connect $s$ to~$t$.
The reduction would certify a non-solution.
This is precisely the configuration the Containment Lemma forbids for shattering cuts, and its breakdown here pinpoints why the $k{+}1$-component threshold is the right one.

\bibliographystyle{alpha}
\bibliography{references}

\end{document}